\documentclass[12pt, journal, onecolumn, twoside]{IEEEtran}
\usepackage{url}
\usepackage[dvips]{color}
\usepackage{algorithm}
\usepackage{algpseudocode}
\usepackage{multirow}
\usepackage{mdframed}
\usepackage{lipsum}
\usepackage{setspace}
\usepackage{graphicx}
\usepackage{graphics}
\usepackage{latexsym}
\usepackage{showidx}
\usepackage{latexsym}
\usepackage{amssymb}
\usepackage{amsfonts}
\usepackage{amsmath}
\usepackage{amssymb}
\usepackage{amsfonts}
\usepackage{amsmath}
\usepackage{xcolor}
\usepackage[normalem]{ulem}
\usepackage{enumitem}
\newtheorem{theorem}{Theorem}

\newtheorem{problem}{Problem}

\newtheorem{example}{Example}

\usepackage{mathtools}
\DeclarePairedDelimiter\floor{\lfloor}{\rfloor}
\DeclarePairedDelimiter{\ceil}{\lceil}{\rceil}
\allowdisplaybreaks
\usepackage[symbol]{footmisc}

\usepackage[tight,footnotesize]{subfigure}

\usepackage{setspace}
\begin{document}
	\title{One-Hop Listening Based ARQs for Low-Latency Communication in Multi-Hop Networks}
	\author{Jaya Goel and J. Harshan\\
		Indian Institute of Technology Delhi, India.\thanks{Parts of this work were presented at IEEE Vehicular Technology Conference (VTC2021)-Spring in April 2021 \cite{our_work_VTC_1}.}}
	\maketitle
	\begin{abstract}
		Inspired by emerging applications in vehicular networks, we address the problem of achieving high-reliability and low-latency communication in multi-hop wireless networks. We propose a new family of Automatic-Repeat-Requests (ARQs) based cooperative strategies wherein high end-to-end reliability is obtained using packet re-transmissions at each hop while the low-latency constraint is met by imposing an upper bound on the total number of packet retransmissions across the network. A hallmark of our strategies is the one-hop listening capability wherein nodes utilize the unused ARQs of their preceding node just by counting the number of failed attempts due to decoding errors. We further extend the idea of one-hop listening to multi-hop listening, wherein a set of consecutive nodes form clusters to utilize the unused ARQs of the preceding nodes, beyond its nearest neighbour, to further improve reliability. Thus, our strategies provide the high-reliability feature with no compromise in the original latency-constraint. For the proposed strategies, we solve non-linear optimization problems on distributing the ARQs across the nodes so as to minimize packet drop probability (PDP) subject to a total number of ARQs in the network. Through extensive theoretical results on PDP and delay profiles, we show that the proposed strategies outperform the best-known strategies in this space.      
	\end{abstract} 
	\begin{small}
		\begin{center}	
		\end{center}
	\end{small}
	\section{Introduction}
	\label{sec:intro}
	\begin{figure}[h!]
		\begin{minipage}{0.45\linewidth}
			\centering
			\includegraphics[width=0.9\textwidth]{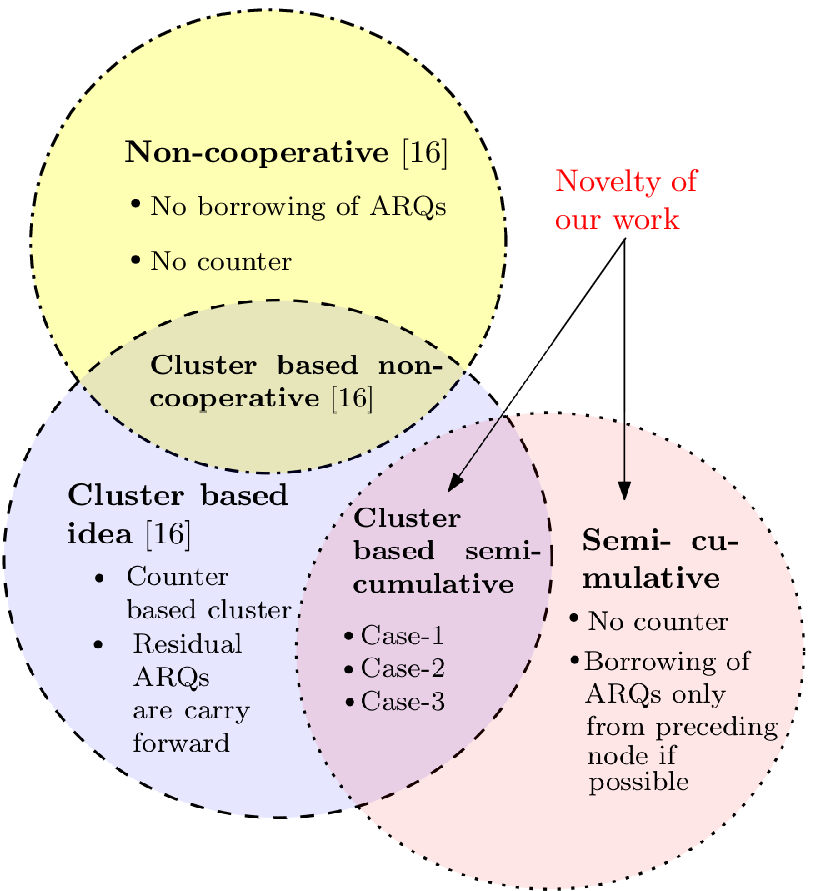}
		\end{minipage}\hfill
		\begin{minipage}{0.55\linewidth}
			\begin{scriptsize}
				\centering
				\begin{tabular}{|c|c|c|}
					\hline
					\textbf{Key points} & \textbf{\cite{our_work_TWC_1}} & \textbf{This work} \\
					\hline 
					\textbf{Without Cluster} &  &  \\
					\hline 
					PDP           & Straight-     & Challenging,  \\
					computation   & forward       & Fibonacci series used \\
					\hline
					Approach  &   Reduction to      & Reduction to solution \\
					&   solution over $\mathbb{R}$   &  over  $(N-2)$ hop network \\
					\hline
					ARQ           &  System of     &  Multi-folding algorithm  \\
					computation   &   linear equations     &  \\
					\hline
					\textbf{With Cluster} &  &  \\
					\hline 
					ARQ distribu-      & Always zero,      & Depends on the cluster\\
					tion at the last   & irrespective of   &  position. Can be zero \\
					node of cluster   & cluster  position.      &  or non-zero.  \\
					\hline
					Use of counter     & In cluster, each node   &  In cluster, the nodes\\
					& uses counter  except    & using counter depends\\
					&  the first node         & on the cluster position\\
					\hline
					Proving      & Based on PDP           &Based on packet  \\
					technique    & at the last node       & survival probability (PSP) \\
					& of the cluster.          & due to memory  \\
					\hline
					Complexity      & Same for any          & Changes with the   \\
					reduction      &  position of           & cluster position. \\
					technique       & the cluster.          &  \\
					\hline
				\end{tabular}
			\end{scriptsize}	
		\end{minipage}
		\caption{\label{fig:venn_diagram}Left-side: Among the existing contributions, \cite{our_work_TWC_1} proposed a non-cooperative strategy (marked in yellow) and a cluster-based idea (marked in blue) along with the non-cooperative strategy. The proposed semi-cumulative strategy (marked in pink) outperforms \cite{our_work_TWC_1} with no additional delay-overheads on the packet. Right-side: Table summarizing the technical differences between \cite{our_work_TWC_1} and this work.}
	\end{figure}
	Multi-hop networks have found promising applications in wireless communication as they enhance the reliability of communication between wireless devices that are either outside of each other's coverage area or unable to establish a communication link due to signal-blockage difficulties \cite{ Zhou2015, Shaikh2018}, \cite{Badarneh2016}. While the problem of designing efficient and reliable protocols for multi-hop networks have been traditional topics of interest, recent advances in the field of vehicle-to-vehicle/infrastructure (V2X) communication have given rise to new requirements such as high-reliability and low-latency on the protocols \cite{Pocovi2018, Ji2018}. Example use-cases include autonomous V2X communication, in which strict deadlines are imposed on the round-trip delay between the vehicles and the infrastructure, after which either the messages become stale or the deadline violation may result in catastrophic consequences \cite{Schulz2017}. Other well-known use-cases of high-reliability and low-latency in multi-hop networks are networks of Unmanned Aerial Vehicles (UAVs) \cite{Changyang_She,Hong_Ren,Chen2018}, in which power-limited UAVs either act as relays in coordinating the movement of autonomous vehicles, or act as airborne base-stations \cite{ Mozaffari_LOS, Hourani_UAV_application, Hourani_UAV_application2, Lin_UAV_application}. Thus, owing to increasing use-cases for achieving high-reliability and low-latency in multi-hop networks, the problem of designing efficient, reliable and importantly \emph{low-latency} protocols, is of utmost importance in the context of next-generation networks. Some of the notable contributions in this field hitherto include  \cite{Changyang_She}, \cite{Hong_Ren}, \cite{Chen2018}, \cite{our_work_VTC_1} and \cite{our_work_TWC_1}. Among them, \cite{our_work_TWC_1} attempted to jointly accomplish high-reliability along with low-latency feature by proposing ARQ based Decode and Forward (DF) strategies  \cite{Wiemann2005}. In such ARQ based DF strategies, a given node is allowed to retransmit the packet a certain number of times when the next node in the chain is unable to decode the packet. In particular, \cite{our_work_TWC_1} takes a two-step approach: (i) a deadline on the end-to-end delay for the packets is imposed depending on the application, and then (ii) using the deadline, a strategy is proposed to forward the packets such that the fraction of packets that reach the destination within the deadline is maximized. Although the ARQ based DF protocols are known to incur additional delay due to the use of ACK/NACK in the reverse channel, they have been shown to offer reduced average end-to-end delay on the packets when compared to fixed block-length coding schemes for a given worst-case deadline constraint \cite[Section IV.D]{our_work_TWC_1}.
	This is because the ARQ based strategies ask for re-transmissions only when the channels are in deep-fade thereby reducing the total number of re-transmissions when the packet travels through the network. As a result, even though the additional overhead for ACK/NACK is added to the delay in each hop, the packets reach the destination within the deadline with high probability since lower number of re-transmissions dominates the overhead introduced by ACK/NACK \cite[Section IV.D]{our_work_TWC_1}. Furthermore, it has been shown that events of deadline violation occur with negligible probability especially when (i) the time/frequency resources dedicated for ACK/NACK communication are negligible compared to that of the payload, and (ii) the frame structure for low-latency communication is such that the ACK/NACK in the reverse link is communicated immediately after decoding the packet. 
		
	In order to motivate this work, we first explain the \emph{worst-case deadline} approach of \cite{our_work_TWC_1} by assuming that the delay overheads from ACK/NACK in the reverse channel are sufficiently small compared to the payload. Suppose that the processing time at each hop is $\tau_{p}$ seconds (which includes packet encoding and decoding time), the delay incurred for packet transmission at each hop is $\tau_{d}$ seconds (which includes the propagation delay and the time-frame of the packet), and the delay incurred because of NACK overhead is $\tau_{NACK}$ (which is the time taken for the transmitter to receive the NACK). Given the stochastic nature of the wireless channel at each link, the total number of packet re-transmissions before the packet reaches the destination is a random variable, denoted by $n$, and as a result, the end-to-end delay between the source and the destination is upper bounded by $n \times(\tau_{p} + \tau_{d} + \tau_{NACK})$ seconds. In particular, when $\tau_{NACK} << \tau_{p} + \tau_{d}$, the end-to-end delay can be approximated as $n \times(\tau_{p} + \tau_{d})$ seconds. Thus, when the packet size and the decoding protocol at each node are established, and when the deadline on end-to-end delay (denoted by $\tau_{total})$ is known, we may impose an upper bound on $n$, provided by $q_{sum} = \floor{\frac{\tau_{total}}{\tau_{p}+\tau_{d}}}$.\footnote{We have assumed equal $\tau_{p}$ and $\tau_{d}$ at each relay node to obtain a relation between the end-to-end delay $\tau_{total}$ and the total number of ARQs $q_{sum}$. Note that equal $\tau_{d}$ holds in practice because the packet length is the same at each hop whereas the propagation delay is usually negligible. However, if $\tau_{p}$ at each relay differs due to heterogeneous architecture for baseband signal processing, then an upper bound on the total number of ARQs can still be obtained by considering the maximum of the processing delays offered by all the relays in the chain.} This implies that $q_{sum}$ captures the maximum number of re-transmissions that can be tolerated over the multi-hop network in order to respect the deadline on the delay. In the event when $\tau_{NACK}$ is not negligible compared to $\tau_{p} + \tau_{d}$, we have the option of either using $q_{sum} = \floor{\frac{\tau_{total}}{\tau_{p}+\tau_{d}}}$ or $q_{sum} = \floor{\frac{\tau_{total}}{\tau_{p}+\tau_{d} + \tau_{NACK}}}$. In the former case, while a larger fraction of packets reach the destination due to higher $q_{sum}$, a non-zero fraction of the packets that reach the destination may arrive after the deadline, thereby violating the latency constraint. However, in the latter case, although a smaller fraction of the packets reach the destination due to lower $q_{sum}$, all of them arrive within the deadline. Thus, with either options for deciding $q_{sum}$, performance degrades as $\tau_{NACK}$ increases. Henceforth, along the similar lines of \cite{our_work_TWC_1}, we use $q_{sum} = \floor{\frac{\tau_{total}}{\tau_{p}+\tau_{d}}}$ assuming that the frame structure and the resources for ACK/NACK communication support the condition $\tau_{NACK} << \tau_{p} + \tau_{d}$.
	
	Once $q_{sum}$ captures the latency-constraint, the subsequent task is to handle the reliability metric by distributing the $q_{sum}$ ARQs across the nodes. With this worst-case deadline approach, several cooperative strategies were proposed in  \cite{our_work_TWC_1} to distribute $q_{sum}$ number of ARQs, for any $q_{sum} \in \mathbb{Z}_{+}$, such that the reliability metric of packet drop probability (PDP) is minimized for any combination of Line-of-Sight (LOS) components of the links. In particular, the cooperative strategies use a counter in the packet so that the unused ARQs by a node can be used by the succeeding nodes in order to further reduce the PDP. Although the cooperative strategies of \cite{our_work_TWC_1} are appealing, the use of counters in the packet contributes to additional communication-overhead in the packet. Therefore, we ask: (i) \emph{Are there cooperative strategies for ARQ based DF protocols that DO NOT use a counter in the packet, and yet utilize the unused ARQs without violating the latency constraint?}, and (ii) \emph{If the use of counter is allowed, are there cooperative strategies that outperform \cite{our_work_TWC_1}?} Towards answering these questions:
		
	
	1) We propose a cooperative ARQ based DF protocol, referred to as the Semi-Cumulative (SC) strategy, wherein every node is aware of the ARQs allotted to its preceding node in the network. As a result, each node can use the unused ARQs of the preceding node, just by counting its number of failed packet transmissions due to decoding errors (i.e., one-hop listening). Unlike the cooperative strategy of \cite{our_work_TWC_1}, the SC strategy does not use a counter in the packet, and yet facilitates efficient utilization of unused ARQs without compromising the latency constraint. Owing to the memory-property of sharing the unused ARQs, we propose a Fibonacci series based method to write the PDP expression as a function of the ARQs. Subsequently, we formulate an optimization problem for computing the optimal ARQ distribution that minimizes the PDP subject to a sum constraint on the total number of ARQs. Furthermore, we present theoretical results to compute near-optimal ARQ distributions using low-complexity methods, and show that the SC strategy outperforms the non-cooperative method with no compromise on the delay constraints.
	
	2) Despite using the SC strategy, there may be residual ARQs at some nodes which go unused. This is because a node cannot listen to the number of incorrect decoding events of all the preceding nodes in the chain. To circumvent this problem, we propose a Cluster-based Semi-Cumulative (CSC) strategy in which a group of consecutive nodes form a cluster, such that every node in the cluster forwards the information on the residual ARQs of its preceding node to the next node in the cluster through a counter in the packet. Therefore, a node in the cluster can make use of the residual ARQs of all the preceding nodes in the cluster. However, the nodes outside the cluster continue to use only the residual ARQs of their immediately preceding node without the need for a counter. This way, we further reduce the PDP of the network from that of the SC strategy without violating the sum constraint. Given the memory property introduced by the residual ARQs, we first provide a method to write the PDP of the CSC strategy, and then formulate an optimization problem to minimize the PDP subject to the sum constraint on the total ARQs. Furthermore, theoretical results on the ARQ distribution within the cluster and outside the cluster are also provided before proposing several low-complexity algorithms to solve the optimization. 
	
	3) Through extensive simulation results, we show that the proposed low-complexity algorithms for the CSC strategy provide near-optimal ARQ distributions in minimizing the PDP. Furthermore, we show that SC strategy and its cluster variant respectively outperform the non-cooperative strategy and its cluster variant \cite{our_work_TWC_1} for a given number of total ARQs. In addition, unlike the cluster based non-cooperative strategy, we show that the performance of the CSC strategy depends on the position of the cluster in the network. This is because the unused ARQs of the last node of the cluster will have to be used by the next node in the chain. We also present simulation results on packet delay profiles to highlight the impact of cooperation on latency performance. We show that with no overhead to access residual ARQs from the packet, the CSC strategies incur a marginal increase in average delay compared to the SC strategy. This is due to more re-transmissions to provide lower PDP, and the need for updating the counter once at each intermediate relay. However, it allows a majority of packets to reach the destination within the deadline. 
	
	In order to implement the proposed one-hop listening based strategies, every node requires the knowledge of the ARQs allotted to itself and its preceding node. One possible strategy to achieve this task is to use a control center that collects the long-term statistics of every node’s channel. Subsequently, the control center can compute the optimal ARQ allocation for each node under a sum constraint on the total number of ARQs, and then distribute the optimal ARQs to each node along with the ARQs allotted to the preceding nodes. This overhead for backhaul coordination is minimal when the long-term statistics of the channels vary slowly over time. As depicted in Fig. \ref{fig:venn_diagram} and explained in the motivation part,  \cite{our_work_TWC_1} is closest to our contributions. 
	A summary of the key differences between our work and \cite{our_work_TWC_1} is also listed on the right side of Fig. \ref{fig:venn_diagram}. Other contributions that use ARQs for optimizing the network performance are \cite{MohitSharma2016}, \cite{CoopARQ_EnergyHarvesting}, \cite{CoopARQ_Diversity}, \cite{Martin_Serror}. None of these contributions address the optimal allocation of ARQs when the total number of ARQs is constrained. A preliminary version of this work is in \cite{our_work_VTC_1}, wherein the SC strategy is proposed. In addition to the contents of \cite{our_work_VTC_1}, this work applies the cluster based ideas on the SC strategy.
	\section{Semi-Cumulative ARQ Strategy for Multi-Hop Networks}
	\label{sec:semiCumm_network_model}
\begin{figure}
		\begin{center}
			\includegraphics[scale=1.2]{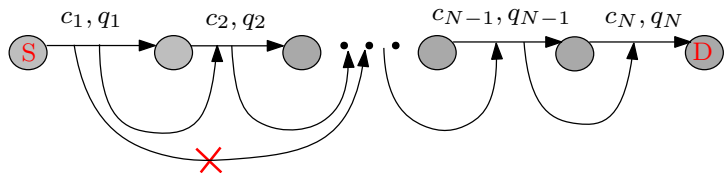}
			\vspace{-0.2cm}
			\caption{\label{Network_model_1}Depiction of an $N$-hop network implementing the SC strategy. The vector $[c_{1}, c_{2}, \ldots, c_{N}]$  denotes the LOS vector of the network, and $q_{i}$, for $1 < i \leq N$, denotes the ARQs allotted to the $i$-th hop. In this model, every node knows the number of ARQs allotted to its preceding node. As a result, it can use the residual ARQs of the preceding node just by listening to the number of failed decoding attempts.}
		\end{center}
	\end{figure}

	Consider an $N$-hop network, as shown in Fig. \ref{Network_model_1}, wherein a source node intends to communicate its messages to a destination through a set of $N-1$ relay nodes that operate using an ARQ based decode and forward (DF) strategy. In this model, the multi-hop network is characterized by the LOS vector $\mathbf{c}= \{c_{1}, c_{2},\ldots,c_{N}\}$ and the ARQ distribution $\mathbf{q}= \{q_{1}, q_{2},\dots,q_{N} \}$, such that $c_{i} \in [0,1]$ represents the LOS component of the fading channel of the $i$-th hop and $q_{i}$ represents the number of re-transmissions allotted to the transmitter of the $i$-th hop, for $1 \leq i \leq N$. Formally, let $\mathcal{S} \subset \mathbb{C}^{K}$ denote the channel code employed at the source node of rate $R$ bits per channel use, i.e., $R = \frac{1}{K}\log_{2}(|\mathcal{S}|)$. Let $\mathbf{x} \in \mathcal{S}$ denote the packet (traditionally referred to as a codeword) transmitted over the multi-hop network such that $\frac{1}{K} \mathbb{E}[|\mathbf{x}|^{2}] = 1$. When $\mathbf{x}$ is transmitted over the $i$-th link, for $1 \leq i \leq N$, the corresponding received signal after $K$ channel uses is given by $\mathbf{y}_{i} = h_{i}\mathbf{x} + \mathbf{n}_{i} \in \mathbb{C}^{K}$, where $h_{i}$ is a quasi-static Ricean fading channel given by $h_{i} = \sqrt{\frac{c_{i}}{2}}(1+\iota)+\sqrt{\frac{(1-c_{i})}{2}}g_{i},$ such that $\iota = \sqrt{-1}$, $g_{i}$ is distributed as $\mathcal{CN}(0,1)$, $\mathbf{n}_{i}$ is the additive white Gaussian noise (AWGN) vector at the receiver of the $i$-th link, distributed $\mathcal{CN}(0,\sigma^{2}\mathbf{I}_{K})$. We assume that the receiver of each link has  perfect knowledge of its channel, however, there is no knowledge about the channel at the transmitter side.
	
	When decoding the packet in the $i$-th hop, if the instantaneous mutual information of the channel is less than the transmission rate $R$, then the receiver will not able to decode the packet correctly. In the regime of asymptotic block-lengths, i.e., $K \rightarrow \infty$, this event completely characterizes the decoding error probability of the $i$-th hop, denoted by $P_{i}$. In particular, with asymptotic block-lengths, $P_{i}$ would be $\mbox{Prob} \Big( R > \log_{2}(1+ |h_{i}|^{2}\alpha ) \Big),$ where $\alpha =  \frac{1}{\sigma^{2}}$ is the average signal-to-noise-ratio (SNR) of the $i$-th link. However, in the regime of non-asymptotic block-lengths, i.e., when $K< \infty$, the corresponding non-asymptotic decoding error probability, as derived in \cite{V_Poor}, can be computed as 
		\begin{equation}
			\label{eq:outage_prob_link}
			P_{i} = \int_\mathbb{R^{+}} Q\left (  \sqrt{\frac{K}{V(\Gamma_{i})}}(C(\Gamma_{i})-R)\right )f_{\Gamma_{i}}(\Gamma_{i})d\Gamma_{i},
		\end{equation}
		where $\Gamma_{i} = |h_{i}|^2\alpha$ is the instantaneous SNR of the wireless channel in the $i$-th hop, $f_{\Gamma_{i}}(.)$ is the probability density function of the instantaneous SNR $\Gamma_{i}$, $C(\Gamma_{i}) = \mbox{log}_2(1 + \Gamma_{i})$ is the Shannon channel capacity, $V(\Gamma_{i}) = \frac{\Gamma_{i}}{2}\frac{\Gamma_{i} + 2}{(\Gamma_{i}+1)^2}\mbox{log}^2_2e$ is the back-off factor for finite block-length, and finally, $Q(x)= \frac{1}{\sqrt{2\pi}}\int_{x}^{\infty}e^{-\frac{u^2}{2}}du$. Note that the expression in \eqref{eq:outage_prob_link}, which is derived using the achievable rates in \cite{polyanskiy}, is applicable for any $K$, and it collapses to the asymptotic outage probability expression as a special case when $K \rightarrow \infty$. 	
	Owing to the above mentioned error events, the transmitter of the $i$-th hop is allotted $q_{i}$ number of re-transmissions in order to successfully forward the packet to the next node in the network. The receiver of the $i$-th hop, for each $i$, uses the Type-1 ARQ model, wherein the received signal $\mathbf{y}_{i}$ corresponding to a failed attempt is not used to decode the subsequent attempts. Despite using this ARQ based DF strategy, if a transmitter is unable to transmit the packet within $q_{i}$ number of ARQs, then the packet is said to be dropped in the network. Since the packet can be dropped at any hop in the network, we use PDP as the reliability metric of interest, which is defined as the fraction of packets that do not reach the destination. 
	
	To achieve higher reliability than the ARQ based DF protocol in \cite{our_work_TWC_1}, we propose the semi-cumulative model, as shown in Fig. \ref{Network_model_1}, wherein every intermediate relay node can use residual ARQs unused by its previous node in the chain. This simple idea stems from the fact that although $q_{i}$ re-transmissions are allotted to a transmitter, the actual number of re-transmissions can be less than $q_{i}$ owing to the stochastic nature of the wireless channel. To facilitate this, we assume that every node has the knowledge of the number of ARQs given to its preceding node in addition to the ARQs allotted to itself. Thus, we call this strategy a ``one-hop listening" strategy. Since a given node can use unused ARQs from its previous node, the total number of ARQs used by it can be more than the number of ARQs allotted to it. As a result, the next node in the chain, despite knowing the number of ARQs allotted to its preceding node, does not know how long to wait for the successful transmission of the packet especially when the preceding node has used all its quota of assigned ARQs. To fix this coordination issue, each intermediate node will have to wait for a fixed amount of time to receive the packet from its previous node beyond which the packet is said to be dropped in the network. Unlike the non-cooperative strategy of \cite{our_work_TWC_1}, in this method, an intermediate node can get more re-transmissions than the number allotted to it just by listening to the number of failed decoding attempts of the preceding node. Note that no counter is needed in the packet, and as a result, the communication-overhead is zero. Although each relay node is allowed  multiple transmissions (including the number of unused ARQs of its preceding node) to communicate the packet to the next node, there is a non-zero probability with which the packet is dropped in the network since the sum of the ARQs allotted to all the nodes in the network is bounded, i.e., $\sum_{i=1}^{N}q_{i} = q_{sum}$. Henceforth, we denote the PDP of the SC ARQ based DF strategy by $pdp_{s, N}$, where $s$ in the subscript highlights the SC strategy, and $N$ denotes the number of hops in the network. Note that $pdp_{s, N}$ is some function of $\{P_{i}, 1 \leq i \leq N\}$ given in \eqref{eq:outage_prob_link}. Thus, to provide reliability along with low-latency constraint, we propose to solve Problem \ref{opt_problem_1}:
	\begin{mdframed}
		\begin{problem}
			\label{opt_problem_1}
			For an $N$-hop network with LOS vector $\mathbf{c}$, a given SNR $\alpha = \frac{1}{\sigma^{2}}$, and a given $q_{sum}$, solve $q_{1}^{*},q_{2}^{*},\ldots q_{N}^{*}=\arg\underset{q_{1},q_{2},\ldots q_{N}}{\text{min}}\ pdp_{s, N}$, $\text{subject to} ~q_{i}\in \{\mathbb{Z}_{+} \cup 0\},\ \forall i, \mbox{~such that ~}\sum_{i = 1}^{N} q_{i} = q_{sum}.$
		\end{problem}
	\end{mdframed}
	\maketitle
	%
	\IEEEpeerreviewmaketitle
	\subsection{PDP Expression of Semi-Cumulative Strategy} 
	\label{sec:analysis}
	Towards solving Problem \ref{opt_problem_1} with low-complexity algorithms, we derive an expression for the PDP of the SC strategy, and then formally prove that the SC strategy outperforms the non-cooperative ARQ strategy in \cite{our_work_TWC_1}. Since the preliminary version of this work in \cite{our_work_VTC_1} already contains the SC strategy, we refer the readers to \cite{our_work_VTC_1} for the proofs.
	\begin{theorem}
		\label{thm1}
		The PDP expression for an $N$-hop SC strategy is given by
		\begin{equation}
			pdp_{s,N} = P_{1}^{q_{1}} + (1-P_{1})P_{2}^{q_{2}}F_{2} + \ldots + \prod_{i = 1}^{N-1}(1-P_{i})P_{N}^{q_{N}}F_{N},
		\end{equation}
		where $F_{j}$, for $2 \leq j \leq N$, is a function of $P_{1}, P_{2}, \ldots, P_{j-1}$ (as given in \eqref{eq:outage_prob_link}) and $q_{1}, q_{2}, \ldots, q_{j-1}$ that can be computed using Fibonacci series.
	\end{theorem}
	\begin{theorem}  
		\label{thm2}
		For a given $\mathbf{q} = [q_{1},q_{2},\dots,q_{N}]$, at high SNR values, the PDP of the SC strategy is upper bounded by the PDP of the non-ARQ strategy.
	\end{theorem}	
	Using $pdp_{s, N}$, we propose low-complexity algorithms to solve Problem \ref{opt_problem_1} in the next section.
	\subsection{Optimal ARQ Distribution of the SC Strategy}
	\label{sec:opt_distribution}
	For an $N$-hop network with $\mathbf{q}= [q_{1},q_{2},\ldots,q_{N-1}, q_{N}]$, suppose that the ARQs for the first $N-2$ hops are fixed, and we are interested in computing the optimal values of $q_{N-1}$ and $q_{N}$ that minimizes the PDP. If we start with $\tilde{\mathbf{q}}= [q_{1},q_{2},\ldots,0, q_{N-1}+q_{N}]$, it may give us a sub-optimal PDP. Therefore, using $\tilde{\mathbf{q}}$, as we keep transferring one ARQ from the last node to the penultimate node, we can expect the PDP to decrease, and then start to increase beyond a certain number of transfers. Towards understanding this transition of PDP, we are interested in understanding the structure of the ARQ distribution when the PDPs of network with $\mathbf{q}= [q_{1},q_{2},\ldots,q_{N-1}, q_{N}]$ and $\mathbf{q}^{'}= [q_{1},q_{2},\ldots,q_{N-1}+1, q_{N}-1]$ are equal. Once we obtain this relation, we can analytically compute the values of $q_{N-1}$ and $q_{N}$ for a given $q_{1}, q_{2}, \ldots, q_{N-2}$, which in turn reduces the search space for computing the optimal ARQ distribution. This result is formally captured in the following theorem.
	\begin{theorem}
		\label{complexity_them_1}
		To find the optimal distribution of ARQs for an $N$-hop network, brute force search can be reduced into brute force search for $(N-2)$-hop network by fixing ARQs $q_{1}, q_{2}, \ldots, q_{N-2}$. 
	\end{theorem}	
	In Theorem \ref{complexity_them_1}, we have proved that the search space for the $N$-hop network can be reduced to the search space of an $(N-2)$-hop network. Henceforth, we refer to this reduction as a one-fold technique. For a large value of $N$, we observe that the one-fold technique may not be feasible to implement in practice. Therefore, we propose low-complexity algorithms to further reduce the search space under the framework of multi-folding algorithms, that are generalizations of the one-fold algorithm.
	
	\section{Low-complexity Algorithms for the SC Strategy}
	\label{sec:proposed_method}
	
	In the proposed multi-folding algorithm, as presented in Algorithm \ref{multi-level-algo}, instead of folding the network once from $N$-hop to $(N-2)$-hop, we fold it multiple times to $(N-4)$-hop, $(N-6)$-hop and so on up to a $2$-hop network or a $1$-hop network depending on whether $N$ is even or odd, respectively. When the network is reduced (or folded) to a $j$-hop network, we need to provide a sum of $\tilde{q}_{sum, j}$ ARQs to it, and it is clear that $\tilde{q}_{sum,j}$ can take all possible values in a range $[j, q_{sum}-(N-j)+1]$. When folding the network up to $j$-hops, for $j \geq 4$ and $j \geq 3$ when $N$ is even and odd, respectively, we fix the ARQs for the first $(j-2)$-hops and then compute $q_{j - 1}$ and $q_{j}$ using Theorem \ref{complexity_them_1} for each value of $\tilde{q}_{sum,j}$. Subsequently, we create a list of ARQ distributions [$q_{1},\ldots,q_{j}$], denoted by $\mathcal{L}_{j}$, by varying the values of $\tilde{q}_{sum,j}$. Following a similar procedure, the candidates of $\mathcal{L}_{j}$ are used to generate $\mathcal{L}_{j+ 2}$ for the $(j + 2)$-hop network by using Theorem \ref{complexity_them_1} for each value of $\tilde{q}_{sum,j + 2}$. This way, a list of ARQ distributions are obtained through $\mathcal{L}_{N}$ for the original $N$-hop network. It is clear that the size of the search space $\mathcal{L}_{N}$ reduces with increase in the number of folds. 

	To further reduce the size of the search space from that of Algorithm \ref{multi-level-algo}, we propose to retain the ARQ distribution that gives us minimum PDP for a given $\tilde{q}_{sum,j}$ from the list $\mathcal{L}_{j}$. This way, only one ARQ distribution survives for a given $\tilde{q}_{sum,j}$, thereby significantly reducing the list size when the algorithm traverses to $\tilde{q}_{sum,N}$. In the process of obtaining $q_{j}$ for each $\tilde{q}_{sum,j}$, we minimized the PDP conditioned on $\tilde{q}_{sum,j}$ and $[q_{1}, q_{2}, \ldots, q_{j-2}]$. However, since the optimal distribution of the folded network may not contribute to the optimal distribution of the original $N$-hop network, we also propose to select the ARQ distribution that is second in the list for that $\tilde{q}_{sum,j}$ and $[q_{1}, q_{2}, \ldots, q_{j-2}]$. In other words, for each $\tilde{q}_{sum,j}$ in $\mathcal{L}_{j}$ we choose the ARQ distribution that minimizes the PDP, and for that selected ARQ distribution, we also pick the ARQ distribution obtained by giving one ARQ from the last node to the penultimate node. Evidently, this technique gives a significantly shorter list compared to the multi-folding approach.  
	\begin{algorithm}
		\caption{\label{algo:sc}Multi-folding algorithm for the SC strategy}
		\label{multi-level-algo} 
		\begin{algorithmic}[1]
			\begin{small}
				\Require  $N$, $q_{sum}$, $\mathbf{P} = [P_{1}, P_{2}, \ldots, P_{N}]$.
				\Ensure $\mathcal{L}_{final} $ \Comment{Stores the list of ARQ distributions in search space}
				\State $\mathcal{L}_{k} = \{\phi\}$ for $k = 1,2,\ldots,N$. 
				\If {$N= odd$} \Comment{Start with fixing $q_{1}$} 
				\State $\mathcal{L}_{1} = \{ [1, q_{sum}-(N-1)+1 ]\}$.
				\State Assign $p=3$.
				\For {$j=p:2:N$}
				\For {$i_{1}=1:|\mathcal{L}_{j-2}|$}
				\State $[q_{1}, \ldots, q_{j-2}] = \mathcal{L}_{j-2}({i_{1}})$
				\State Compute $q_{j}$ using $[q_{1}, \ldots, q_{j-2}]$ by applying Theorem \ref{complexity_them_1}
				\For {$\tilde{q}_{sum,j} = j: (q_{sum}-(N-j)+1)$}.
				\State Compute $q_{j-1} = \tilde{q}_{sum,j} - \sum_{t=1, t\neq j-1}^{j}q_{t}$.
				\State Insert $[\mathcal{L}_{j-2}(i_{1})||q_{j-1}||q_{j}]$ to $\mathcal{L}_{j}$ only if $q_{j-1} \geq 0$.
				\EndFor
				\EndFor
				\EndFor
				\State $\mathcal{L}_{final}= \mathcal{L}_{N}$.
				\ElsIf{$N = even$} \Comment{Start with fixing $q_{1}$ and $q_{2}$.}
				\State $\mathcal{L}_{2} = \{\{q_{1},q_{2}\} \in \mathbb{Z}_{+}^{2} | q_{1}+q_{2} \in [2, q_{sum}-(N-2)+1 ]\}$
				\State Assign $p=4$, and repeat steps from line number $5$ to $15$.
				\EndIf
			\end{small}
		\end{algorithmic}
	\end{algorithm}
\vspace{-0.5cm}
	\subsection{Simulation Results and Complexity Analysis}
	\label{sec:Sims}
\begin{figure*}[]
		\begin{center}
			\includegraphics[scale=0.48]{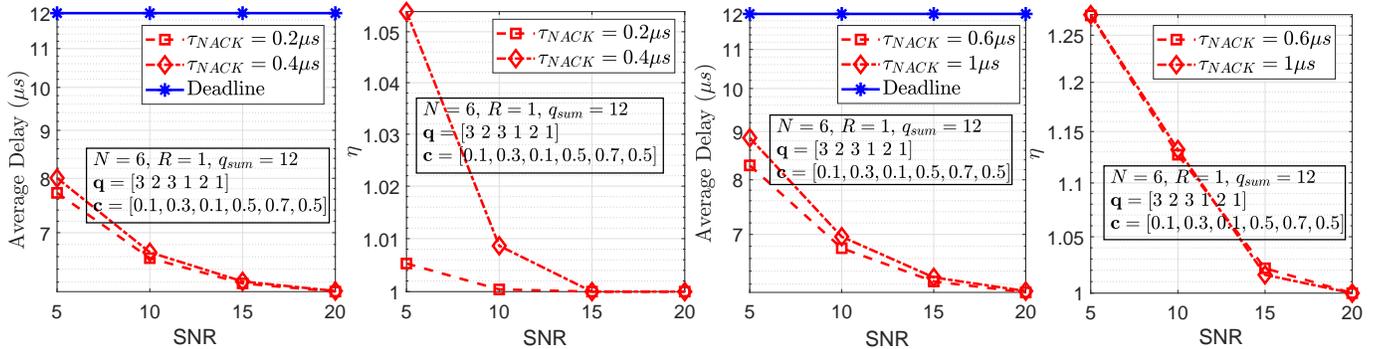}
			\vspace{-1cm}\caption{\label{delay_analysis_sc_strategy} Variation of average delay on the packets and the deadline violation parameter ($\eta$) for various $\tau_{NACK}$ when using the SC strategy.}
		\end{center}
	\end{figure*}

	In the first part, we show that the packets of the ARQ based SC strategy that reach the destination arrive within the given deadline constraint with a high probability, provided the delay overheads from ACK/NACK are sufficiently small. To generate the results, $q_{sum}$ is obtained as $\floor{\frac{\tau_{total}}{\tau_{p}+\tau_{d}}}$ without considering the resources for ACK/NACK in the reverse channel, where $\tau_{total}, \tau_{d}$ and $\tau_{p}$ are as defined in Section \ref{sec:intro}. Subsequently, we introduce different resolution of delays from NACK, say $\tau_{NACK}$ time units, and then study its impact on the end-to-end delay on the packets. Assuming $\tau_{p} + \tau_{d} = 1$ microsecond, we set the deadline for  end-to-end packet delay as $q_{sum}$ microseconds. Then, by sending an ensemble of $10^6 $ packets to the destination through the SC strategy, we compute the following metrics when $\tau_{NACK} \in \{0.2, 0.4, 0.6, 1\}$ microseconds: (i) the fraction of packets that were dropped in the network (denoted by $W_{Drop}$) due to insufficient ARQs at the intermediate nodes, (ii) the fraction of packets that reach the destination after the deadline (denoted by $W_{Deadline}$), and finally, (iii) the average end-to-end delay on the packets. These metrics are plotted in Fig. \ref{delay_analysis_sc_strategy} for various values of $\mbox{SNR}$ at a specific value of $N$ and the LOS vector $\mathbf{c}$. The plots suggest that the average delay is significantly lower than that of the deadline especially when $\tau_{NACK}$ is small, owing to the opportunistic nature of ARQ strategies. However, as $\tau_{NACK}$ increases, the average delay is pushed closer to the deadline. Furthermore, to capture the behaviour of deadline violations due to higher $\tau_{NACK}$, in Fig. \ref{delay_analysis_sc_strategy}, we also plot $\eta = \frac{W_{Drop} + W_{Deadline}}{W_{Drop}}$. The plots confirm that when $\tau_{NACK}$ is sufficiently small compared to $\tau_{p} + \tau_{d}$ (see $\tau_{NACK} = 0.2 \mu s$ at \mbox{SNR} = 15, 20 dB), the packets that reach the destination arrive within the deadline with an overwhelming probability as $\eta = 1$ at those values. 
	
\begin{figure*}[hbt!]
		\begin{center}
			\includegraphics[scale=0.46]{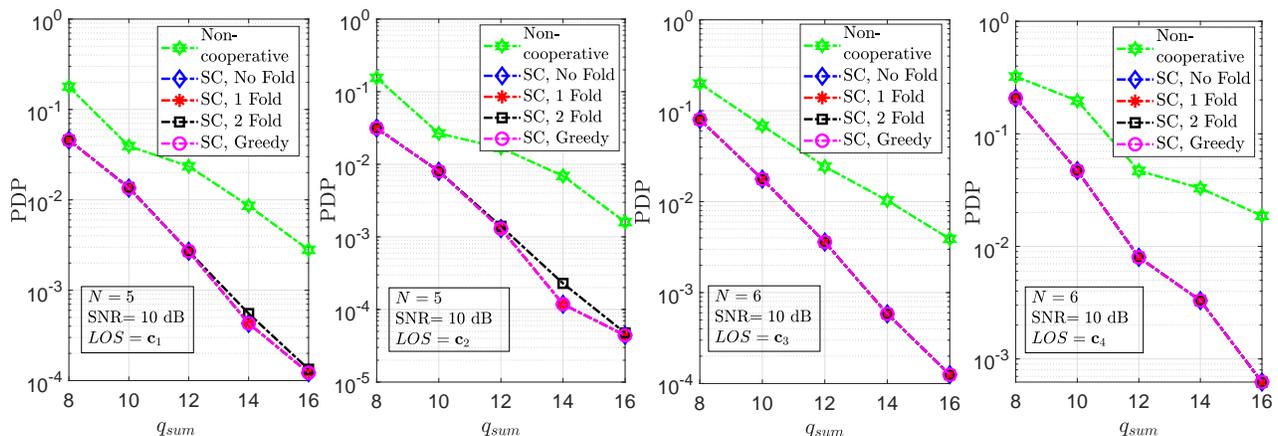}
			\vspace{-0.7cm}
			\caption{\label{pdp_1_N5_6_pdp_qSum_10}PDP plots when using an SC strategy with exhaustive search (no fold), 1-fold, 2-fold and greedy strategies with $\mathbf{c}_{1}=[0.1,0.3,0.1,0.5,0.2]$, $\mathbf{c}_{2}=[0.5,0.5,0.5,0.5,0.5]$, $\mathbf{c}_{3}=[0.9,0.2,0.4,0.7,0.1,0.5]$ and $\mathbf{c}_{4}=[0.3,0.3,0.3,0.3,0.3,0.3]$ at SNR = $10$ dB and rate $R=1$.}
		\end{center}
	\end{figure*}

	In the rest of the section, we present simulation results to analyse the PDP of the SC strategy for various values of $N, q_{sum}$, and LOS vectors. Henceforth, to generate the simulation results for a given LOS vector $\mathbf{c}$ and $\mbox{SNR}$, we use the saddle-point approximation in \cite[Theorem 2]{V_Poor} on \eqref{eq:outage_prob_link} to compute $\{P_{i}, 1 \leq i \leq N\}$. As emphasized in \cite[Section V]{V_Poor}, these approximations are tight for Ricean channels when $R > 0.5$ and when the block-length $K$ is in few hundreds. Therefore, for the proposed approximation to be valid, we use the block-length $K = 500$ in this simulation setup. In general, when the saddle-point approximation in \cite[Theorem 2]{V_Poor} is not tight, $\{P_{i}, 1 \leq i \leq N\}$ in \eqref{eq:outage_prob_link} must be computed using numerical methods. First, in Fig. \ref{pdp_1_N5_6_pdp_qSum_10}, we present simulation results to compare the PDP of the SC strategy with that of the non-cooperative strategy \cite{our_work_TWC_1}. Although we have proved the dominance of our strategy theoretically, the plots confirm  that the PDP of the SC strategy outperforms the PDP of the non-cooperative strategy with no increase in the communication-overhead on the packet. Furthermore, to showcase the benefits of using the multi-fold algorithm and the greedy algorithm, we plot the minimum PDP offered by these algorithms for $N=5$ and $N=6$ in Fig. \ref{pdp_1_N5_6_pdp_qSum_10}. The plots confirm that while the multi-fold algorithm provides near-optimal ARQ distribution, the greedy algorithm is successful in offering the optimal ARQ distributions
	
\begin{figure*}[hbt!]
		\begin{center}
			\includegraphics[scale=0.42]{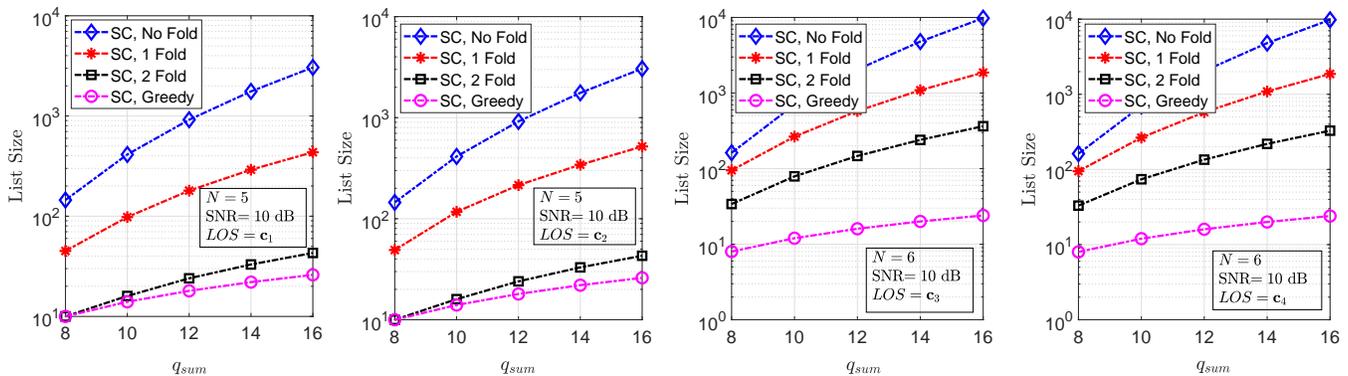}
			\vspace{-0.95cm}
			\caption{\label{list_1_N5_6_pdp_qSum_10}List sizes for an SC strategy with exhaustive search, 1-fold, 2-fold and greedy strategies with $\mathbf{c}_{1}=[0.1,0.3,0.1,0.5,0.2]$, $\mathbf{c}_{2}=[0.5,0.5,0.5,0.5,0.5]$, $\mathbf{c}_{3}=[0.9,0.2,0.4,0.7,0.1,0.5]$ and $\mathbf{c}_{4}=[0.3,0.3,0.3,0.3,0.3,0.3]$ at SNR$=10$ dB and rate $R=1$. }
		\end{center}
	\end{figure*}

	In terms of complexity, for an $N$-hop network, the size of the search space for the SC strategy is upper bounded by $\binom{q_{sum}+N-1}{N-1}$. However, with the multi-fold algorithm, we have shown that the search space can be reduced. To showcase the reduction, we plot the size of the search space ($\mathcal{L}_{N}$) of the multi-fold algorithm for $N=5$ and $N=6$. For these cases, since we can fold the network at most twice, we have shown the results for both one-fold and two-fold cases. The simulation results, as shown in Fig. \ref{list_1_N5_6_pdp_qSum_10}, display significant reduction in the list size as we move to one-fold and two-fold. Finally, the plots also show that the list size of the greedy algorithm is shorter than the multi-fold case, and it is, therefore, amenable to implementation in practice.
	\vspace{-0.5cm}
	
	Although the above presented results showcase the reduction in the overall list size for computing the minimum PDP, they do not capture the number of computations at the destination in order to arrive at the final lists. If we include the computations required to apply the results of Theorem \ref{complexity_them_1} at each level of multi-folding, it is clear that the greedy algorithm offers minimum complexity owing to fewer surviving distributions at each level. From the above results, we conclude that the proposed SC strategy can be preferred to the non-cooperative strategy. In the next section, we explore methods to further improve the reliability of the SC strategy by including the counters in the packet. 
	\section{Cluster Based Semi-Cumulative Strategy}
	\label{sec:cs_semi_cumm}
	In the SC strategy, the benefit of cooperation is limited if an intermediate node uses more ARQs than the number allotted to it. For instance, in a $3$-hop network with ARQ distribution $[q_{1}, q_{2}, q_{3}] = [4, 3, 5]$, suppose that the first hop consumes $2$ attempts and the second hop consumes $4$ attempts by utilizing one residual ARQ from the first hop. Although one residual ARQ is still unused from the first two hops, the third hop cannot utilize this because the second hop has used more ARQs than its allotted quota. On the other hand, if the transmitter of the third hop had the knowledge of residual ARQs entering the transmitter of the second hop, then it would have used that one unused ARQs. Thus, to take advantage of the residual ARQs of the preceding nodes, we require a counter in the packet that would be updated with the residual ARQs at each hop. Formally, the set of consecutive nodes in the network that use a counter to share the residual ARQs in the packet is referred as a cluster \cite{our_work_TWC_1}. To explain the cluster-based idea, with $q_{1}$ denoting the number of ARQs allotted to the first hop, let the first node in the cluster make $q_{1}-r_{1}$ number of attempts to successfully transmit the packet to the second node in the chain, for some $0 \leq r_{1} \leq q_{1} - 1$. After that, when the second node receives the packet successfully, it updates the counter with a number equal to the sum of ARQs allotted to itself and the residual ARQs coming to the previous hop, i.e., $q_{2}+r_{1}$ ARQs, and then transmits the packet to the next node. If the second node of the cluster uses $q_{2} + r_{1} - r_{2}$ attempts, then the third node updates the counter with $q_{3}+r_{2}$ ARQs before transmitting the packet. This way, each receiving node in the cluster updates the counter only once and recovers the total number of unused ARQs by the previous nodes. 
	\begin{figure*}[]
		\begin{center}
			\includegraphics[scale=1.0]{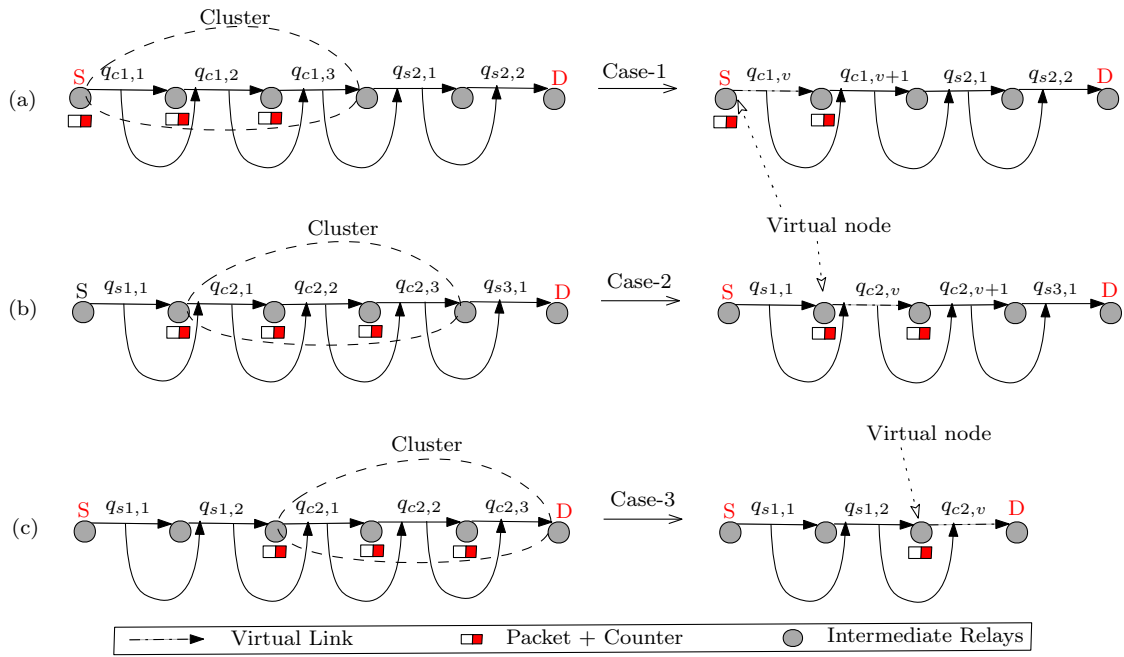}
			\vspace{-0.2cm}
			\caption{\label{Network_model_2} An illustrative example of a $5$-hop network where a cluster is formed by grouping 3 consecutive nodes: (i) the cluster at the beginning (Case-1), (ii)  the cluster at an intermediate position (Case-2), and (iii) the cluster at the end (Case-3).}
		\end{center}
	\end{figure*}
	
	Using the above idea, we propose a Cluster based Semi-Cumulative (CSC) strategy on an $N$-hop network wherein we make a group of nodes that acts as a cluster, as exemplified in Fig. \ref{Network_model_2}. As the grouping of nodes can be done anywhere in the network, we propose three cases, namely, Case-1: cluster placed at the beginning by grouping a set of first few nodes. The network is made up of two portions, a cluster portion followed by a semi-cumulative portion (see (a) in Fig. \ref{Network_model_2}). Case-2: cluster placed at an intermediate position. The $N$-hop network is made up of three portions, a semi-cumulative portion followed by a cluster, which in-turn is followed by a semi-cumulative portion (see (b) in Fig. \ref{Network_model_2}). Case-3: cluster placed at the end by grouping a set of last few nodes in the network. The network is made up of two portions, a semi-cumulative portion followed by a cluster (see (c) in Fig. \ref{Network_model_2}). Overall, the nodes inside the cluster can utilize the unused ARQs of all the preceding nodes in the cluster, whereas the nodes in the semi-cumulative portion(s) utilize the unused ARQs of its preceding node only. Henceforth, a multi-hop network employing the CSC strategy is referred to as the CSC network. 
	
	Similar to the SC strategy, we are interested in computing the optimal ARQ distribution on the $N$-hop CSC network such that its PDP is minimized for a given $q_{sum}$. Let $N_{su}$ and $N_{sw}$ represent the hop sizes of the semi-cumulative portions, and $N_{cy}$ represent the hop size of the cluster. Here the first subscript indicates the type of the sub-network, and the subscripts $u \in \{1\}$ and $y \in \{1, 2\}$ and $w \in \{2, 3\}$ are jointly used to represent their placement in the $N$-hop network. Consequently, we have $N_{su} + N_{cy} + N_{sw} = N$. In particular, the valid combinations of $u, y, w$ that capture the three cases of the CSC network are Case-1: $y = 1$ along with $w = 2$ implying that a cluster of size $N_{c1}$ is followed by a semi-cumulative network of size $N_{s2}$. Case-2: $y = 2$ along with $u = 1$ and $w = 3$ implying that a cluster of size $N_{c2}$ is between two semi-cumulative networks of size $N_{s1}$ and $N_{s3}$. Finally, Case-3: $u = 1$ and $y = 2$ implying that a cluster of size $N_{c2}$ follows a semi-cumulative network of size $N_{s1}$. Furthermore, let the ARQ distribution on the nodes in the cluster be denoted by $\mathbf{q}_{cy} = [q_{cy,1}, q_{cy,2}, \ldots, q_{cy,N_{cy}}]$, and the ARQ distribution on the nodes of the semi-cumulative portions be $\mathbf{q}_{su} = [q_{su,1}, q_{su,2}, \ldots, q_{su,N_{su}}]$ and $\mathbf{q}_{sw} = [q_{sw,1}, q_{sw,2}, \ldots, q_{sw,N_{sw}}]$. Given that a sum constraint is imposed on the ARQs, we have $\sum_{k=1}^{N_{su}}q_{su,k} + \sum_{k=1}^{N_{cy}}q_{cy,k} +\sum_{k=1}^{N_{sw}}q_{sw,k} = q_{sum}$ as long as the combinations of $u, y, w$ are valid. Thus, in contrast to the usual notation on the ARQ vector $\mathbf{q} = [q_{1}, q_{2}, \ldots, q_{N}]$, we use $\mathbf{q} = [\underbrace{q_{su,1},q_{su,2},\dots,q_{su,N_{su}}}_{\mathbf{q}_{su}} \underbrace{q_{cy,1},q_{cy,2},\ldots q_{cy,N_{cy}}}_{\mathbf{q}_{cy}}, \underbrace{q_{sw,1},q_{sw,2},\dots,q_{sw,N_{sw}}}_{\mathbf{q}_{sw}}]$, for valid combinations of $u, y, w$. Similarly, we use $\mathbf{P} = [P_{su,1},P_{su,2},\dots,P_{su,N_{su}} P_{cy,1}, P_{cy,2},\ldots P_{cy,N_{cy}}, P_{sw,1}, P_{sw,2},\dots, P_{sw,N_{sw}}]$ instead of $\mathbf{P} = [P_{1}, P_{2}, \ldots, P_{N}]$, to highlight the association of the outage probabilities to either the semi-cumulative portion or the cluster. With the above notations on a CSC network, the optimization problem for ARQ distribution is given in Problem \ref{opt_problem_2} by encompassing all the three cases. As presented in Problem \ref{opt_problem_2}, $pdp_{cs, N}$ represents the PDP at the destination wherein the subscript $cs$ highlights the CSC network.
	\begin{mdframed}
		\begin{problem}
			\label{opt_problem_2}
			For an $N$-hop CSC network with a given LOS vector $\mathbf{c}$, $N_{cy}$, $N_{su}, N_{sw}$, SNR $\alpha = \frac{1}{\sigma^{2}}$, and $q_{sum}$, solve $\arg\underset{\mathbf{q}}{\text{min}}\ pdp_{cs, N}$, $\text{subject to} ~\mathbf{q} \in \{0 \cup \mathbb{Z}_{+}\}^{N}$ such that $\sum_{k = 1}^{N_{su}} q_{su,k} + \sum_{k = 1}^{N_{cy}} q_{cy,k} + \sum_{k = 1}^{N_{sw}} q_{sw,k} = q_{sum}$ for all valid combinations of $u, y, w$.
		\end{problem}
	\end{mdframed}

Towards solving the above problem, the following questions must be answered, (i) How to write the closed-form expression for $pdp_{cs, N}$ in Case-1, Case-2 and Case-3, (ii) Once the expression for $pdp_{cs, N}$ is written, how to allocate the ARQs to the nodes that are part of the cluster, and those that are outside the cluster. A particularly interesting question under the question in (ii) is ``Does the optimal ARQ distribution on the nodes in the cluster depend on the position of the cluster? To justify the relevance of this question, note that in both Case-1 and Case-2, the node that follows the cluster can use the residual ARQs of the last node of the cluster by listening to its number of failed transmissions. In contrast, in Case-3, there is no node after the cluster that requires unused ARQs. Thus, whether or not non-zero ARQs must be allotted to the last node of the cluster depends on the position of the cluster. Henceforth, in the rest of this section, we follow a $3$-step approach to solve Problem \ref{opt_problem_2}: \textbf{Step 1}: For a given $q_{sum}$, $\mathbf{q}_{su}$ $\mathbf{q}_{sw}$, characterize the structure of $\mathbf{q}_{cy}$ that minimizes the PDP. \textbf{Step 2}: Using the results of \textbf{Step 1}, form a virtual semi-cumulative network, and minimize its PDP. \textbf{Step 3}: Apply the structure of $\mathbf{q}_{cy}$ obtained in \textbf{Step 1} on the solution of \textbf{Step 2}.
	
As part of $\textbf{Step 1}$, we propose to solve Problem \ref{opt_problem_3} that addresses the optimal ARQ distribution within the cluster conditioned on the ARQs allotted to the semi-cumulative network(s). In contrast to Problem \ref{opt_problem_2}, Problem \ref{opt_problem_3} addresses to maximize the Packet Survival Probability (PSP) at the last node of the cluster for any given residual ARQs. Formally, the PSP at the $N_{cy}$-th node of the cluster is denoted by $psp_{cy,N_{cy}}(r_{cy, N_{cy}-1})$, where $r_{cy, N_{cy}-1}$ denotes the number of residual ARQs coming from the $(N_{cy}-1)$-th node of the cluster. We remark that it is imperative to take the PSP approach due to the presence of a semi-cumulative network after the cluster. 
	\begin{mdframed}
		\begin{problem}
			\label{opt_problem_3}
			For an $N$-hop CSC network with a given LOS vector $\mathbf{c}$, SNR $\alpha = \frac{1}{\sigma^{2}}$, $\mathbf{q}_{su}$, $\mathbf{q}_{sw}$ and $q_{sum}$, solve $\arg\underset{q_{cy,1},q_{cy,2},\ldots q_{cy,N_{cy}-1}}{\text{max}}\ psp_{cy,N_{cy}}(r_{cy, N_{cy}-1}), \forall r_{cy, N_{cy}-1}$, where $[q_{cy,1},q_{cy,2},\ldots q_{cy,N_{cy}-1}] \in \{0 \cup \mathbb{Z}_{+}\}^{N_{cy} - 1}$, such that $\sum_{k = 1}^{N_{su}} q_{su,k} + \sum_{k = 1}^{N_{cy}} q_{cy,k} + \sum_{k = 1}^{N_{sw}} q_{sw,k} = q_{sum}$ for all valid $u, y, w$.
		\end{problem}
	\end{mdframed}
	\subsection{Theoretical Results for Cluster based Semi-Cumulative Strategy} 
	First, we consider a CSC network with either Case-1 or Case-2. For a given $\mathbf{q}_{su}$ and $\mathbf{q}_{sw}$, the following theorem proves that except the last node of the cluster, all the ARQs on the nodes inside the cluster must be transferred to the first node of the cluster in order to maximize the PSP at the last node of the cluster.  
	\begin{theorem}
		\label{psp_maximization_case12}
		In Case-1 and Case-2, giving all the ARQs of the first $N_{cy}-1$ nodes of the cluster to the first node of the cluster maximizes the PSP at the $N_{cy}$-th hop of the cluster for any given residual ARQs $r_{cy,(N_{cy}-1)}$ at the $N_{cy}$-th hop.
	\end{theorem}
	\begin{IEEEproof}
		We divide the proof into two parts depending on the placement of the cluster in the network.
		
		\noindent \textbf{Case I} : With the cluster placed at the beginning, we have $N_{c1} + N_{s2}=N$. Let $\mathbf{q}_{c1}= \{q_{c1,1},q_{c1,2}, \ldots, q_{c1,N_{c1}}\}$ and $\mathbf{q}_{s2}= \{q_{s2,1},q_{s2,2}, \ldots, q_{s2,N_{s2}}\}$ represent the ARQ distribution of the networks $c1$ and $s2$, respectively. Therefore, the overall ARQ distribution of the $N$-hop network satisfies the constraint $\sum_{k_{1} = 1}^{N_{c1}} q_{c1,k_{1}} + \sum_{k_{2} = 1}^{N_{s2}} q_{s2,k_{2}} = q_{sum}$. We prove the theorem using the induction method. For the initialization step, let $N_{c1}=3$ and $\mathbf{q}_{c1}= \{q_{c1,1}, q_{c1,2}, q_{c1,3}\}$. Since PSP is the metric of interest, let the residual ARQ arriving at the $3^{rd}$ hop be $r_{c1,2}$, where the range of $r_{c1,2}$ is $[0,q_{c1,1}+q_{c1,2}-2]$. We prove that $\{q_{c1,1}+q_{c1,2}, 0\}$ maximizes the PSP at the $3^{rd}$ hop of the cluster for any residual ARQ. For the ARQ distribution $\{q_{c1,1}, q_{c1,2}\}$, let the consumption profile of ARQs at the first two hops of the cluster be $\tilde{\mathbf{q}}_{c1}= \{\tilde{q}_{c1,1}, \tilde{q}_{c1,2}\}$ such that  $\tilde{q}_{c1,1} \leq q_{c1,1} $ and $q_{c1,1}+q_{c1,2}= \tilde{q}_{c1,1}+ \tilde{q}_{c1,2}+r_{c1,2}$. In order to result in $r_{c1,2}$ residual ARQs, the PSP at the $3^{rd}$ hop of the cluster is $psp_{c1,3}(r_{c1,2}) = (1-P_{c1,1})(1-P_{c1,2}) (\sum_{i=1}^{q_{c1,1}} P_{c1,1}^{i-1}P_{c1,2}^{q_{c1,1}+q_{c1,2}-r_{c1,2}-i-1})$. Note that when the exponent term of $P_{c1,2}$ goes negative, we discard the corresponding terms from PSP expression as those terms are invalid. Similarly, with ARQ distribution $\{q_{c1,1}+q_{c1,2}, 0\}$, the PSP at the $3^{rd}$ hop with $r_{c1,2}$ residual ARQs is $psp_{c1,3}^{'}(r_{c1,2}) = (1-P_{c1,1})(1-P_{c1,2}) (\sum_{i=1}^{q_{c1,1}+q_{c1,2}} P_{c1,1}^{i-1}P_{c1,2}^{q_{c1,1}+q_{c1,2}-r_{c1,2}-i-1})$, 
		and similar to $psp_{c1,3}$, we discard the terms with negative exponent on $P_{c1,2}$. After solving the difference of $psp_{c1,3}^{'}(r_{c1,2})$ and $psp_{c1,3}(r_{c1,2})$, we get $psp_{c1,3}^{'}(r_{c1,2})- psp_{c1,3}(r_{c1,2}) = (1-P_{c1,1})(1-P_{c1,2})(\sum_{i=1}^{q_{c1,2}} P_{c1,1}^{i-1}\\P_{c1,2}^{q_{c1,1}+q_{c1,2}-r_{c1,2}-i-1})$. It can be observed that $psp_{c1,3}^{'}(r_{c1,2}) \geq psp_{c1,3}(r_{c1,2})$ (because $q_{c1,2} \geq 0$) where equality holds when $q_{c1,2}=0 \ \text{or} \ 1$. Therefore, the initialization step is proved. Now, we assume that the result is also true for $N_{c1}=t+1$ for any $t \geq 3$. As a consequence, the optimal ARQ distribution which maximizes the PSP at the $(t+1)$-th hop for any given $r_{c1, t}$ is $\{q_{c1,1}+q_{c1,2}+\ldots+q_{c1,t}, 0, \dots,0 \}$. Now, we have to prove that the same result is true for $N_{c1}= t+2$.
		
		Let $r_{c1,(t+1)}$ be the number of residual ARQs at $(t+2)$-th node, where the range of $r_{c1,(t+1)}$ is $[0, \sum_{i=1}^{t+1} q_{c1,i}-(t+1)]$. Let $psp_{c1,(t+2)}(r_{c1,t+1})$ and $psp_{c1,t+1}(r_{c1,t})$ be the PSP at the $(t+2)$ and $(t+1)$ nodes, respectively. 
		Conditioned on a given $q_{c1,t+1}$, we want to write $psp_{c1,t+2}(r_{c1,t+1})$ in terms of $psp_{c1,t+1}(r_{c1,t})$. This means for a given $r_{c1,t}$, the $(t+1)$-th node must make $(q_{c1,t+1}+r_{c1,t}-r_{c1,t+1})$ attempts out of which one is successful and the others are unsuccessful. The PSP at $(t+2)$-th node is
		\begin{equation}
			\label{psp_t+2}	
			psp_{c1,t+2}(r_{c1,t+1})= (1-P_{c1,t+1}) \sum_{r_{c1,t}} psp_{c1,t+1}(r_{c1,t})P_{c1,t+1}^{q_{c1,t+1}+r_{c1,t}-r_{c1,t+1}-1},
		\end{equation}
		where $(1-P_{c1,t+1})P_{c1,t+1}^{q_{c1,t+1}+r_{c1,t}-r_{c1,t+1}-1}$ represents the probability that $q_{c1,t+1}+r_{c1,t}-r_{c1,t+1}$ ARQs are consumed at $(t+1)$-th node. Note that once $r_{c1,t}$ is fixed, the second term in the summation of (\ref{psp_t+2}) is fixed. This implies that to maximize $psp_{c1,t+2}(r_{c1,t+1})$ for a given $q_{c1,t+1}$, we need to maximize $psp_{c1,t+1}(r_{c1,t})$ for every $r_{c1,t}$ in the valid range. From the induction step, we have already assumed that the ARQ distribution $\{q_{c1,1}+q_{c1,2}+\ldots+q_{c1,t},0,\ldots,0\}$ maximizes the PSP for any residue at the $(t+1)$-th hop. Therefore, by invoking the result from $(t+1)$-hop network, the optimal ARQ distribution for the $(t+2)$-hop network conditioned on $q_{c1,t+1}$ is of the form $\mathbf{q}_{c1,t+1}$ = $\{q_{c1,1}+q_{c1,2}+ \ldots+q_{c1,t}, 0, \ldots,0, q_{c1,t+1}\}$. As the last step of this proof, we need to show that $q_{c1,t+1}$ must be transferred to the first node of the cluster in order to maximise the PSP for any residue at the $(t+2)$-th hop. Using proof by contradiction, let us assume that $q_{c1,t+1} > 1$ maximizes the PSP for any residue at the $(t+2)$-th hop. In that case, let us focus on the ARQ consumption profile $\tilde{\mathbf{q}}_{c1,t+1}= \{\tilde{q}_{c1,1}, \tilde{q}_{c1,2}, \ldots, \tilde{q}_{c1,t+1}\}$ of the first $t + 1$ nodes of the cluster that results in $r_{c1,t+1}=0$ at node $t + 2$. We immediately note that when $q_{t+1}>1$ in $\mathbf{q}_{c1,t+1}$, it is not possible to have ARQ consumption of the form $\{q_{c1,1}+ q_{c1,2} + \ldots + q_{c1,t}-(t-1)+1,1,\ldots, 1, q_{c1,t+1}-1\}$ because a node cannot borrow from its succeeding node. Therefore, the mass point value on the above consumption profile is $0$. On the other hand, if $q_{c1,t+1}=0 \ \text{or} \ 1$ in $\mathbf{q}_{c1,t+1}$, we can obtain every possible ARQ consumption profile, and hence in this case, we have a non-zero mass point value against each ARQ consumption profile. This is a contradiction as it results in higher value of $psp_{c1,t+2}(r_{c1,t+1} = 0)$ compared to that when $q_{t+1}>1$ in $\mathbf{q}_{c1,t+1}$. Thus, the optimal ARQ distribution that maximizes PSP for every residue at the $(t+2)$-th hop is $\{q_{c1,1}+q_{c1,2}+\ldots+q_{c1,t+1},0, \ldots, 0\}$. Although the ARQ on the $(t+1)$-th node can be either $0$ or $1$, we have used $0$ in this proof.\\
		\noindent  \textbf{Case II} : In this case, the cluster is placed in between two semi-cumulative networks such that $N= N_{s1}+N_{c2}+N_{s3}$. As a consequence, the first node of the cluster can make use of the residual ARQs from its previous node in addition to the ARQs allotted to it. Let $r_{s1,N_{s1}} \in [0,q_{s1,N_{s1}}-1]$ be the residual ARQs from the last node of the $s1$ network. Therefore, the first node of the cluster can use ARQs in the range $[q_{c2,1} , q_{c2,1}+ r_{s1,N_{s1}} ]$. Since a semi-cumulative network precedes the cluster, the PSP at the first and the last nodes of the cluster are given in \eqref{conditional_psp_first_node_cluster} and \eqref{conditional_psp_cluster}, respectively.  -
		\begin{equation}
			\label{conditional_psp_first_node_cluster}
			psp_{c2,1}(r_{s1,N_{s1}}) = (1-P_{s1,N_{s1}}^{q_{s1,N_{s1}}}) \sum_{r_{s1,N_{s1}-1}}psp_{s1,N_{s1}}( r_{s1,N_{s1}-1})P_{s1,N_{s1}}^{q_{s1,N_{s1}}-1-r_{s1,N_{s1}}}.
		\end{equation}           
		\begin{equation}
			\label{conditional_psp_cluster}	
			psp_{c2,N_{c2}}(r_{c2,N_{c2}-1}) = psp_{s1,N_{s1}}(r_{s1,N_{s1}-1})psp_{c2,N_{c2}}(r_{c2,N_{c2}-1}|r_{s1,N_{s1}-1}, q_{c2,1}, \dots, q_{c2,N_{c2}-1}).
		\end{equation} 
		The LHS of \eqref{conditional_psp_first_node_cluster} is fixed since the ARQ distribution on the semi-cumulative part $s1$ is fixed. However, the LHS of \eqref{conditional_psp_cluster} depends on the ARQs allotted to the first $N_{c2} - 1$ nodes of the cluster as well as how the residual ARQ entering the first node of the cluster is used. It can be observed that in order to maximize the LHS of \eqref{conditional_psp_cluster} we need to maximize the second term of (\ref{conditional_psp_cluster}) because $psp_{s1,N_{s1}}(r_{s1,N_{s1}-1})$ is constant. By using Case I, we can maximize the second term by transferring all the ARQs of the first $N_{c2} -1$ nodes along with the residual ARQs from $s1$ to the first node of the cluster. This completes the proof.   
	\end{IEEEproof}
	
	The following theorem shows that the above discussed ARQ distribution within the cluster minimizes the average PDP at the destination for a given ARQ allocation on the semi-cumulative networks.
	\begin{theorem}
		\label{prop:PDP_minimization}
		In Case-1 and Case-2, for a given $\mathbf{q}_{su}$ and $\mathbf{q}_{sw}$, by maximizing the PSP at the $N_{cy}$-th hop of the cluster for any residual ARQs, we  minimize the average PDP at the destination.
	\end{theorem}
	\begin{IEEEproof}
		First, we consider Case-1 where the cluster of size $N_{c1}$ is followed by a semi-cumulative network of size $N_{s2}$. For this case, we have already proved that the ARQ distribution $\{q_{c1,1}+q_{c1,2}+\ldots+q_{c1,N_{c1}-1}, 0, \ldots, 0 \}$ maximizes the PSP at $N_{c1}$-th hop of the cluster for every $r_{c1,N_{c1}-1}$. Furthermore, let $q_{c1,N_{c1}}$ be the number of ARQs given to the last node of the cluster, and therefore, the $N_{c1}$-th node can use upto $q_{c1,N_{c1}}+ r_{c1,N_{c1}-1}$ ARQs. If $N_{c1}$-th node uses more than $q_{c1,N_{c1}}$ attempts (because of residual ARQs as indicated in the packet), then the first node of the $s2$ network cannot get ARQ benefits. However, if the $N_{c1}$-node uses fewer than $q_{c1,N_{c1}}$ attempts, then the first node of the $s2$ network can borrow the residual ARQs unused by the $N_{c1}$-th node. Formally, we can write the PSP at the first node of $s2$ network with respect to the PSP at $N_{c1}$-th hop as $psp_{s2,1}(r_{c1,N_{c1}}) = (1-P_{c1,N_{c1}})\sum_{r_{c1,N_{c1}-1}}psp_{c1,N_{c1}}(r_{c1,N_{c1}-1})P_{c1,N_{c1}}^{q_{c1,N_{c1}}-1+r_{c1,N_{c1}-1}-r_{c1,N_{c1}}}$, where $r_{c1,N_{c1}-1}$ and $r_{c1,N_{c1}}$ represent the number of residual ARQs arriving at $N_{c1}$-th node and first node of $s2$ network, respectively. Similarly, the PSP at the second node of the $s2$ network and at the destination are respectively given by $$psp_{s2,2}(r_{s2,1}) = (1-P_{s2,1})\sum_{r_{c1,N_{c1}}}psp_{s2,1}(r_{c1,N_{c1}})P_{s2,1}^{q_{s2,1}-1+r_{c1,N_{c1}}-r_{s2,1}},$$ $$psp_{s2,D}(r_{s2,N_{s2}}) = (1-P_{s2,N_{s2}})\sum_{r_{s2,N_{s2}-1}}psp_{s2,N_{s2}}( r_{s2,N_{s2}-1})P_{s2,N_{s2}}^{q_{s2,N_{s2}}-1+r_{s2,N_{s2}-1}-r_{s2,N_{s2}}},$$ where $D$ represents the destination of the $N$-hop network (i.e. the $N_{s2}$-th hop). It must be observed that since $psp_{c1,N_{c1}}(r_{c1,N_{c1}-1})$ is maximized for each $r_{c1,N_{c1}-1}$, the term $psp_{s2,1}(r_{c1,N_{c1}})$ is maximized for each $r_{c1,N_{c1}}$. Furthermore, by repeating the steps in the similar way, we maximize the PSP at the destination for every $r_{s2,N_{s2}}$. As a result, the average PSP is maximized at the destination, which in turn minimizes the average PDP at the destination. This completes the proof when the cluster is placed at the beginning. We can use a similar approach to prove the statement of the theorem for Case-2.
	\end{IEEEproof}
	\begin{theorem}
		In Case-3, for a given $\mathbf{q}_{su}$ and $q_{sum}$, giving all the ARQs of the $N_{cy}$ nodes of the cluster to the first node of the cluster minimizes the PDP at the destination
	\end{theorem}
	\begin{IEEEproof}
		The proof is along the similar lines of Theorem \ref{psp_maximization_case12} and Theorem \ref{prop:PDP_minimization} with the exception that the ARQs allotted to the last node of the cluster must also be transferred to the first node of the cluster; this is because there is no semi-cumulative network following the cluster in Case-3.
	\end{IEEEproof}
	
	For a given CSC network, once the ARQs on the $N$ nodes is known, we have so far completed \textbf{Step 1} in our approach. In order to complete \textbf{Step 2}, it is important to write the expression for the PDP of the network. Given that it is challenging to write the PDP expression owing to the memory property in the strategy, we provide a set of rules to write the PDP expression for any $N$. Similar to the PDP expression for the SC strategy, we continue to make use of binary sequences, however, without using the structure of Fibonacci series owing to the presence of the cluster in the network. For the ease of explaining our procedure, if the cluster is placed other than the last position in the $N$-hop network, we split the cluster into two parts. The first part of the cluster is the group of first $N_{cy} -1$ nodes, henceforth referred to as the virtual node called node $v$. Similarly, the last node of the cluster is referred to as node $v+1$. On the other hand, when the cluster is placed at the last position of the network, all the $N_{cy}$ nodes of the cluster are treated as node $v$. Therefore, we replace the cluster by either two virtual hops or one virtual hop depending on its location in the network. Consequently, we will replace the physical $N$-hop network by a virtual $\tilde{N}$-hop network, where $\tilde{N} = N - (N_{c1} - 1) + 1$, $\tilde{N} = N - (N_{c2} - 1) + 1$, and $\tilde{N} = N - (N_{c2}) + 1$, for Case-1, Case-2, and Case-3, respectively. On this $\tilde{N}$-hop network, the effective ARQ vector is given by $\mathbf{q} = [q_{su, 1}, q_{su, 2}, \ldots, q_{su, N_{su}}, q_{v}, q_{v+1}, q_{sw, 1}, q_{sw, 2}, \ldots, q_{sw, N_{sw}}] \in \{0 \cup \mathbb{Z}_{+}\}^{\tilde{N}}$, where $q_{v} = \sum_{i = 1}^{N_{cy}-1} q_{cy, i}$, and $q_{v+1}= q_{cy, N_{cy}}$ for Case-1 and Case-2, whereas $q_{v} = \sum_{i = 1}^{N_{cy}} q_{cy, i}$ for Case-3. The procedure for writing the PDP expression for this virtual $\tilde{N}$-hop network is explained in the next theorem.
	\begin{theorem}
		\label{thm_closed_form_CSC}
		After replacing the cluster by node $v$ and node $v+1$ in an $N$-hop CSC network, the PDP expression for the $\tilde{N}$-hop network can be written in closed-form. 
	\end{theorem}
	\begin{IEEEproof}
		For the virtual $\tilde{N}$-hop network, we can write the PDP expression at the destination as $pdp_{cs, \tilde{N}} = pdp_{cs,1h}+ pdp_{cs,2h} + \ldots + pdp_{cs,\tilde{N}h}$, where $pdp_{cs,kh}$, for $1 \leq k \leq \tilde{N}$, denotes the probability that the packet is dropped at the $k$-th hop. To write the expression of $pdp_{cs,kh}$, for each $k \in \{1,2,\ldots, \tilde{N}\}$, we recommend using a set of $k$-length binary sequences to characterize the packet surviving event till the $k$-th hop. Similar to Theorem \ref{thm1}, while writing the expression for $pdp_{cs,kh}$, a bit $`0'$ in the $m$-th position of the sequence, for $1 \leq m < k$, indicates that the $m$-th node forwards the packet to its next node without borrowing the residual ARQs from its preceding node. Similarly,  bit $`1'$ indicates that the $m$-th node forwards the packet after borrowing the residual ARQs from its preceding node. For a given $k$, we discard the invalid sequences that do not follow the rules of CSC network These rules are: (a) sequences starting with $1$ at the MSB must be discarded since the first node is the source, (b) sequence containing all $0$ is invalid, (c) in a sequence, two consecutive $'1'$ can only occur at the positions of $v$ and $v+1$ nodes of the cluster, because node $v+1$ of the cluster can borrow from node $v$ irrespective of whether node $v$ node has borrowed from its preceding node or not. This event occurs when the cluster is placed other than last position in the $N$-hop network. Therefore, a $k$-length sequence is invalid if it contains consecutive ones at any positions except for the above mentioned case, and (d) the last bit of the sequence can be either $`0'$ or $`1'$ as it represents whether the packet is dropped at the $k$-th node with no borrowing or borrowing of residual ARQs from the $(k-1)$-th node, respectively. First, we collect the set of valid $k$-length sequences, and then based on whether $k$ is odd or even, we propose a procedure to use the $k$-length sequences to construct $pdp_{cs, kh}$. We divide the $k$-length sequence into the chunks of bits from the left to the right such that first chunk contains only the MSB bit and the subsequent bits are divided into the chunks of two bits. If $k$ is odd, we end up with a chunk of two bits at the end, otherwise, we end up with a chunk of one bit. To map the bits to corresponding PDP terms, we parse the chunks of the sequence from left to right, and then replace the chunks with the terms as listed in Table.\ref{tab:table_2}. Among the terms in the table, the notation $psp_{v}(\cdot)$ represents the probability that packet reaches node $v+1$ after consuming as many ARQs in the argument by the virtual node. Similarly, $pdp_{v}(\cdot)$ represents the probability that the packet is dropped by the virtual node despite consuming as many ARQs in the argument. In this process, it is important to identify the locations of node $v$ and node $v+1$ for every $k$. This way, each valid binary sequence is written as the product of terms of the form $\beta_{\gamma_{1},\gamma_{2} I}$, $\beta_{\gamma_{1},\gamma_{2}I}$, $\beta_{\gamma_{1},\gamma_{2}E}$, and $\beta_{\gamma_{1},\gamma_{2}E}$ as given in Table.\ref{tab:table_2}, for valid combinations of $\gamma_{1} \in \{0, 1,01, 10, 11\}$ and $\gamma_{2} \in \{s,v, v+1, ss, sv, v(v+1), (v+1)s\}$. Once a $k$-length sequence is replaced by the corresponding expression, we add the expressions of all the valid $k$-length sequences to obtain $B_{k}$. Finally, using $B_{k}$, we obtain (i) $pdp_{cs, kh} = \prod_{i = 1}^{k-1}(1-P_{s1, i})(P_{s1, k})^{q_{s1, k}}B_{k}$, when $k \leq N_{s1}$, (ii) $pdp_{cs, kh} = \prod_{i = 1}^{k-1}(1-P_{s1, i})B_{k}$, when $k = N_{s1} + 1$, (iii) $pdp_{cs, kh} = \prod_{i = 1}^{k-2}(1-P_{s1, i})(P_{v+1})^{q_{v+1}}B_{k}$, when $k=N_{s1}+2$ (referred to as a special case in Table. \ref{tab:table_2}) and (iv) $pdp_{cs, kh} = (1 - P_{v+1}) \prod_{i = 1}^{N_{s1}}(1-P_{s1, i})\prod_{j = 1}^{t-1}(1-P_{s2, j})(P_{s2, t})^{q_{s2, t}}B_{k}$, when $k \geq N_{s1} + 2 + t$, where $1 \leq t \leq N_{s2}$ for all above cases. Overall, we get $pdp_{cs,\tilde{N}} = \sum_{i=1}^{\tilde{N}}pdp_{cs,ih}$.     
	\end{IEEEproof}
	\begin{table*}[b!]
		\begin{scriptsize}
			\begin{center}
				\caption{Binary sequence based PDP expression for $\tilde{N}$-hop CSC network}
				\label{tab:table_2}
				\begin{tabular}{|c|c|c|c|c|c|c|c|} 
					\hline
					\textbf{Nodes} & \textbf{Chunk} & \textbf{Chunks} & $\mathbf{\beta}$s &  \textbf{Expression} \\
					& \textbf{size} &  & \textbf{terms} &\\
					\hline
					\textbf{First} & & & &\\
					\textbf{position} & & & &\\
					\hline
					s (MSB)  & 1 & $0$ & $\beta_{0,sI}$ & $\sum_{i=1}^{q_{s1,1}}P_{s1,1}^{q_{s1,1}-i}$, \\
					$v$ (MSB)  & 1 & $0$ & $\beta_{0,vI}$ & $\sum_{i=1}^{\bar{q}_{v}}psp_{v}(q_{v}-i+1)$, \text{where} \\ 
					& & & &  $\bar{q}_{v} = q_{v} - (N_{c1}-2)$. \\
					\hline
					\textbf{Middle } & & & & For $(s, s)$ combinations, the indices\\
					\textbf{positions} & & & & $m, m -1, m+1$ refer to positions of the nodes\\
					\textbf{$m < \tilde{N}$} & & & & inside the semi-cumulative portions\\
					\hline
					s, s & 2 & $00$ & $\beta_{00,ssI}$  & $\big( \sum_{i=1}^{q_{sx,(m-1)}}P_{sx,m-1}^{q_{sx, m-1}-i} \big)\big(\sum_{j=1}^{q_{sx, m}}P_{sx, m}^{q_{sx,m}-j} \big)$, for $x \in \{u, w\}$.  \\
					s, s & 2 & $01$ & $\beta_{01,ssI}$ & $\big( \sum_{i=1}^{q_{sx,m-1}}P_{sx,m-1}^{q_{sx,m-1}-i} \big)\big( \sum_{j=0}^{i-2}P_{sx,m}^{q_{sx,m}+j} \big)$.  \\
					s, s & 2 & $10$ & $\beta_{10,ssI}$ & $\big( \sum_{j=0}^{i-2}P_{sx,m}^{q_{sx,m}+j} \big)\big( \sum_{k=1}^{q_{sx,(m+1)}}P_{sx,(m+1)}^{q_{sx,(m+1)}-k} \big)$, \\
					& & &  & where $i$ is residual ARQs from previous node. \\
					\hline
					s, $v$ & 2 & $00$ & $\beta_{00,svI}$  & $\big( \sum_{i=1}^{q_{s1,N_{s1}}}P_{s1, N_{s1}}^{q_{s1, N_{s1}}-i} \big)\big( \sum_{j=1}^{\bar{q}_{v}}psp_{v}({q_{v}-j+1}) \big)$.  \\
					s, $v$ & 2 & $01$ & $\beta_{01,svI}$ & $\big( \sum_{i=1}^{q_{s1,N_{s1}}}P_{s1,N_{s1}}^{q_{s1,N_{s1}}-i} \big)\big(\sum_{j=1}^{i-1}psp_{v}({q_{v}+i-1-j+1}) \big)$.  \\
					s, $v$ & 2 & $10$ & $\beta_{10,svI}$ &$\big( \sum_{j=0}^{i-2}P_{s1,N_{s1}}^{q_{s1,N_{s1}}+j} \big)\big(\sum_{k=1}^{\bar{q}_{v}}psp_{v}({q_{v}-k+1})\big)$, \\
					& & &  & where $i$ is residual ARQs from previous node. \\
					\hline
					$v+1$, s & 2 & $00$ & $\beta_{00,(v+1)sI}$ & $\big( \sum_{i=1}^{q_{v+1}}P_{v+1}^{q_{v+1}-i} \big)\big( \sum_{i=1}^{q_{sw,1}}P_{sw,1}^{q_{sw,1}-i} \big)$. \\ 
					$v+1$, s & 2 & $01$ & $\beta_{01,(v+1)sI}$ & $\big( \sum_{i=1}^{q_{v+1}}P_{v+1}^{q_{cy,v+1}-i} \big)\big( \sum_{j=0}^{i-2}P_{sw,1}^{q_{sw,1}+j} \big) $.\\ 
					$v+1$, s & 2 & $10$ & $\beta_{10,(v+1)sI}$ & $\big( \sum_{j=0}^{i-1}P_{v+1}^{q_{v+1}+j} \big)\big( \sum_{k=1}^{q_{sw,1}}P_{sw,1}^{q_{sw,1}-k} \big)$.\\ 
					\hline
					$v$, $v+1$ & 2 & $00$ & $\beta_{00,v(v+1)I}$ & $\big(\sum_{i=1}^{\bar{q}_{v}}psp_{v}(q_{v}-i+1) \big)\big( \sum_{j=1}^{q_{v+1}}P_{v+1}^{q_{v+1}-j} \big)$. \\ 
					$v$, $v+1$ & 2 & $01$ & $\beta_{01,v(v+1)I}$ & $\big( \sum_{i=1}^{\bar{q}_{v}}psp_{v}(q_{v}-i+1) \big)\big(\sum_{j=0}^{i-1}P_{v+1}^{q_{v+1} + j} \big)$.  \\ 
					$v$, $v+1$ & 2 & $10$ &  $\beta_{10,v(v+1)I}$ & $\big( \sum_{j=1}^{i-1}psp_{v}(q_{v}+i-1-j+1) \big)\big( \sum_{k=1}^{q_{v+1}}P_{v+1}^{q_{v+1}-k} \big)$,  \\
					& & &  & where $i$ is the residual ARQs from previous node. \\
					$v$, $v+1$ & 2 & $11$ & $\beta_{11,v(v+1)I}$ & $\big( \sum_{j=1}^{i-1}psp_{v}(q_{v}+i-1-j+1) \big)\big( \sum_{k=0}^{j-1}P_{v+1}^{q_{v+1}+k} \big)$,  \\
					& & &  & where $i$ is the residual ARQs from previous node. \\
					\hline
					\textbf{Last} & & & &\\
					\textbf{positions} & & & &\\
					\hline
					s, s & 2 & $01$ & $\beta_{01,ssE}$ & $\sum_{i=1}^{q_{sw,N_{sw}}-1}P_{sw,N_{sw}-1}^{q_{sw,N_{sw}-1}-i}P_{sw,N_{sw}}^{i-1}$, \\
					s, s & 2 & $10$ & $\beta_{10,ssE}$ & $\sum_{j=0}^{i-2}P_{s1,N_{s1}-1}^{q_{s1,N_{sw}-1}+j}$ \\
					\hline
					s, $v$ & 2 & $01$ & $\beta_{01,svE}$ & $\sum_{i=1}^{q_{s1,N_{s1}}}P_{s1,N_{s1}}^{q_{s1,N_{s1}}-i}PDP_{v}(q_{v}+i-1)$,  \\
					s, $v$ & 2 & $10$ & $\beta_{10,svE}$ & $\sum_{j=0}^{i-2}P_{s1, N_{s1}}^{q_{s1, N_{s1}}+j}PDP_{v}(q_{v})$ \\
					\hline
					$v+1$, s & 2 & $01$ &  $\beta_{01,(v+1)sE}$  & $\sum_{i=1}^{q_{v+1}}P_{v+1}^{q_{v+1}-i}P_{sw,1}^{i-1}$, \\
					$v+1$, s & 2 & $10$ & $\beta_{10,(v+1)sE}$ & $\sum_{j=0}^{i-2}P_{v+1}^{q_{v+1}+j}$  \\
					\hline
					$v$, $v+1^{*}$ & 2 & $01$ &  $\beta_{01,v(v+1)E}$  & $\big( \sum_{i=1}^{\bar{q}_{v}}psp_{v}(q_{v}-i+1) \big)P_{v+1}^{i-1}$.  \\ 
					$v$, $v+1^{*}$ & 2 & $11$ &  $\beta_{11,v(v+1E}$  & $\big( \sum_{j=1}^{i-1}psp_{v}(q_{v}+i-1-j+1) \big)P_{v+1}^{j-1}$,  \\
					* Special&  & &  & where $i$ is the residual ARQs from previous node. \\
					case&  & &  & \\
					\hline
					& & & & \\
					$s$ (LSB) & 1 & $0$ & $\beta_{0, sE}$ & 1, \\
					$s$ (LSB) & 1 & $1$ & $\beta_{1,sE}$ & $P_{sw,N_{sw}}^{i-1}$ where $i$ is the residual ARQs from the previous node.  \\ 
					$v$ (LSB) & 1 & $1$ & $\beta_{1, vE}$ & $PDP_{v}(q_{v}+i-1) $ where $i$ is the residual ARQs from the previous node.  \\ 
					\hline
				\end{tabular}
			\end{center}
		\end{scriptsize}
	\end{table*}
	\begin{example}
		We pick the case wherein the cluster is followed by a semi-cumulative network with $N_{c1}=3$ and $N_{s2}=2$. The network is shown in Fig. \ref{Network_model_2}(a) and the ARQ distribution is given by $\textbf{q} = [q_{v},q_{v+1}, q_{s2,1},q_{s2,2}]$, where $q_{v}$ is the number of ARQs given to the first $N_{c1}-1$ hops of the cluster and $q_{v+1}$ is the number of ARQs given to the last node of the cluster. Therefore, a $5$-hop network can be visualized as a $4$-hop virtual network. To write the expression for $pdp_{cs,4h}$, the valid $4$-length sequences are $\{0101, 0010, 1101\}$, where we map $'0101'$ to $(\sum_{i=1}^{q_{v}}psp_{v}(q_{v}-i-1)\sum_{j=1}^{i-1} P_{v+1}^{q_{v+1}+j})(\sum_{i=1}^{q_{s2,1}}P_{s2,1}^{q_{s2,1}-i}P_{s2,1}^{q_{s2,1}+i-1})$, $'0010'$ to $(\sum_{i=1}^{q_{v}}psp_{v}(q_{v}-i)) (\sum_{i=1}^{q_{v+1}} P_{v+1}^{q_{v+1}-i}P_{s2,1}^{q_{s2,1}+i-1})( P_{s2,2}^{q_{s2,2}})$ and $'1101'$ to $(\sum_{i=1}^{q_{v}}psp_{v}(q_{v}-i))\\ (\sum_{i=1}^{q_{v+1}} P_{v+1}^{q_{v+1}-i})(\sum_{i=1}^{q_{s2,1}}P_{s2,1}^{q_{s2,1}-i}P_{s2,2}^{q_{s2,2+i-1}})$. Similarly, we can write $pdp_{cs,kh}$ for $k = 1, 2, 3$ by using valid $k$-length binary sequences, and then write the overall PDP as $pdp_{cs, 4} = pdp_{cs,1h}+ pdp_{cs,2h}+ pdp_{cs,3h} + pdp_{cs,4h}$.
	\end{example}
	
	From the above results, we have shown that the PDP for the CSC strategy can be written using the ARQ vector is $\mathbf{q} = [q_{su, 1}, \ldots, q_{su, N_{su}}, q_{v}, q_{v+1}, q_{sw, 1}, \ldots, q_{sw, N_{sw}}] \in \{0 \cup \mathbb{Z}_{+}\}^{\tilde{N}}$, where $q_{v} = \sum_{i = 1}^{N_{cy}-1} q_{cy, i}$, and $q_{v+1}= q_{cy, N_{cy}}$. Since $\mathbf{q}$ satisfies the sum constraint $\left(\sum_{k=1}^{N_{su}}q_{su,k}\right) + q_{v} + q_{v+1} + \left(\sum_{k=1}^{N_{sw}}q_{sw,k}\right) = q_{sum}$, it is straightforward to compute the optimal ARQ distribution $[q_{su, 1}^{*}, \ldots, q_{su, N_{su}}^{*}, q_{v}^{*}, q_{v+1}^{*}, q_{sw, 1}^{*}, \ldots, q_{sw, N_{sw}}^{*}]$ through exhaustive search. Subsequently, we can complete \textbf{Step 3} by obtaining the optimal ARQ distribution of the $N$-hop network as $q_{cy, 1}^{*} = q_{v}^{*}$, $q_{cy, N_{cy}}^{*} = q_{v+1}^{*}$ and $q_{cy, j}^{*} = 0$ for $1 < j < N_{cy}$. However, since exhaustive search is not practical, we present theoretical results on complexity reduction to compute the optimal ARQ distribution for all the three cases under \textbf{Step 2}. 
	\subsection{Complexity Reduction for Case-1 and Case-2}	
	In the following theorem, we show that the optimal ARQ distribution of the $\tilde{N}$-hop network can be computed by using the search space for the first $\tilde{N}-2$ values of $\mathbf{q}$.
	\begin{theorem}\label{thm:complexity_reduction_cs_case1_2}
		For Case-1 and Case-2, to find the optimal ARQ distribution, the brute force search of an $\tilde{N}$-hop network can be reduced to the brute force search for $(\tilde{N}-2)$-hop network. 	
	\end{theorem}	
	\begin{IEEEproof}
		Let the ARQ distribution of the $\tilde{N}$-hop network be $\tilde{\mathbf{q}}_{cs,\tilde{N}}=[\tilde{q}_{cs,1}, \tilde{q}_{cs,2}, \ldots, \tilde{q}_{cs,\tilde{N}-1}, \tilde{q}_{cs,\tilde{N}}]$, where either $\tilde{q}_{cs,\tilde{N}-1} = q_{sw, N_{sw}-1}, \tilde{q}_{cs,\tilde{N}} = q_{sw, N_{sw}}$, or $\tilde{q}_{cs,\tilde{N}-1} = q_{v+1}, \tilde{q}_{cs,\tilde{N}} = q_{sw, 1}$. Similarly, let the outage probability vector be $[\tilde{P}_{cs, 1}, \tilde{P}_{cs, 2}.\ldots, \tilde{P}_{cs, \tilde{N}}]$ = $[P_{su, 1}, \ldots, P_{su, N_{su}}, P_{v}, P_{v+1}, P_{sw, 1}, \ldots, P_{sw, N_{sw}}]$, where $P_{v} = psp_{v}(\{q_{cy, 1}, \ldots, q_{cy, N_{cy}-1}\})$ is as defined in Theorem \ref{thm_closed_form_CSC}, $P_{v+1} = P_{cy, N_{cy}}$. Our approach for the proof is to fix the ARQs for the first $\tilde{N}-2$ nodes, and then transfer the ARQs from the last node to the penultimate node until the PDP is minimized. When we give one ARQ from last node to the penultimate node, we obtain $\tilde{\mathbf{q}}^{'}_{cs,\tilde{N}} = [\tilde{q}_{cs,1}, \tilde{q}_{cs,2}, \ldots,\tilde{q}_{cs,\tilde{N-1}}+1, \tilde{q}_{cs,\tilde{N}}-1]$. Now, the PDP expressions with $\tilde{\mathbf{q}}_{cs,\tilde{N}}$ and $\tilde{\mathbf{q}}^{'}_{cs,\tilde{N}}$, respectively, can be written as $pdp_{cs,\tilde{N}} = pdp_{cs,\tilde{N}-2} + pdp_{cs,(\tilde{N}-1)h}+ pdp_{cs,\tilde{N}h}$, $pdp_{cs,\tilde{N}}^{'} = pdp_{cs,\tilde{N}-2}^{'} + pdp_{cs,(\tilde{N}-1)h}^{'}+ pdp_{cs,\tilde{N}h}^{'},$ where the individual expressions are the probabilities that the packet is dropped at the intermediate links. Also, it can be noted that the last two nodes must be either only semi-cumulative nodes or a $v+1$ node that is followed by a semi-cumulative node. It is straightforward to note that $pdp_{cs,jh} = pdp_{cs,jh}^{'}$ for $1\leq j \leq \tilde{N}-2$ since the first $\tilde{N}-2$ terms are the same in $\tilde{\mathbf{q}}_{cs,\tilde{N}}$ and $\tilde{\mathbf{q}}^{'}_{cs,\tilde{N}}$, when cluster is placed other than the last position. Therefore, similar to Theorem \ref{complexity_them_1}, on equating $pdp_{cs,\tilde{N}}= pdp_{cs,\tilde{N}}^{'}$, we get $pdp_{cs,(N-1)h}- pdp_{cs,(\tilde{N}-1)h}^{'} =-(pdp_{cs,\tilde{N}h} - pdp_{cs,\tilde{N}h}^{'}),$ where we can write $pdp_{cs,(\tilde{N}-1)h}^{'}= \tilde{P}_{cs,\tilde{N}-1}\left(pdp_{cs,(\tilde{N}-1)h}\right)$ because at the $(\tilde{N}-1)$-th hop, every term of $B_{\tilde{N}-1}^{'}$ gets multiplied by $\tilde{P}_{cs,\tilde{N}-1}$ (outage probability of the $(\tilde{N}-1)$-th hop) since one ARQ has been transferred from the $\tilde{N}$-th hop. Hence, we can write $pdp_{cs,(\tilde{N}-1)h}(1-\tilde{P}_{cs,\tilde{N}-1}) = -(pdp_{cs,\tilde{N}h} - pdp_{cs,\tilde{N}h}^{'}).$ On expanding the above equation and including $(1-\tilde{P}_{cs,\tilde{N}-1})$ in the product loop, we can write
		\begin{small}
			\begin{eqnarray}
				\label{eq:indept_binary}
				& & \bigg(\frac{B_{\tilde{N}-1}}{\prod_{i=1, i \neq j, s.t. \tilde{P}_{cs, j} = P_{v}}^{\tilde{N}-2}\tilde{P}_{cs,i}^{q_{cs,i}}} \bigg)  = \tilde{P}_{cs,\tilde{N}}^{\tilde{q}_{cs,\tilde{N}}}\frac{(\tilde{P}_{cs,\tilde{N}}^{-1}B_{\tilde{N}}^{'}-B_{\tilde{N}})}{\prod_{i=1, i \neq j, s.t. \tilde{P}_{cs, j} = P_{v}}^{\tilde{N}-1}\tilde{P}_{csi}^{\tilde{q}_{cs,i}}}.
			\end{eqnarray}
		\end{small}
		\noindent where $B_{\tilde{N}}$ and $B_{\tilde{N}}^{'}$ are the terms obtained using the binary sequence representation corresponding to  $pdp_{cs,\tilde{N}h}$ and $pdp_{cs,\tilde{N}h}^{'}$, respectively. In the rest of the proof, we will show that $\frac{\tilde{P}_{cs,\tilde{N}}^{-1}B_{\tilde{N}}^{'}-B_{\tilde{N}}}{\tilde{P}_{cs,\tilde{N}-1}^{\tilde{q}_{cs,\tilde{N}-1}}}$ does not contain $\tilde{q}_{cs,\tilde{N}-1}$ in it. Towards that direction, note that both $B_{\tilde{N}}^{'}$ and $B_{\tilde{N}}$ contain the same number of terms in their expansion using binary sequences, however, with the difference that the terms $\tilde{q}_{cs,\tilde{N}}$ and $\tilde{q}_{cs,\tilde{N}-1}$ in $B_{\tilde{N}}$ appear as $\tilde{q}_{cs,\tilde{N}} - 1$ and $\tilde{q}_{cs,\tilde{N}-1} + 1$ in $B_{\tilde{N}}^{'}$, respectively. When constructing $B_{\tilde{N}}^{'}$ and $B_{\tilde{N}}$ using binary sequences of length $\tilde{N}$, we partition the terms of $B_{\tilde{N}}^{'}$ and $B_{\tilde{N}}$ into two categories, namely: the sequences that end with `01' and sequences that end with `10'. This is because the states of the nodes before the last two digits are the same for both $B_{\tilde{N}}^{'}$ and $B_{\tilde{N}}$. As a result, for the sequences that end with `01', we can take the term $\beta_{01,\gamma_{2}E}$, for $\gamma_{2} \in \{(v+1)s, ss\}$ at the locations $\tilde{N} - 1$ and $\tilde{N}$, common, and only focus on its effect in $\frac{B_{\tilde{N}}^{'}-\tilde{P}_{cs,\tilde{N}}B_{\tilde{N}}}{\tilde{P}_{cs,\tilde{N}-1}^{\tilde{q}_{cs,\tilde{N}-1}}}$. Similarly, for the sequences that end with `10', we can take the term $\beta_{10,\gamma_{2}I}$, for $\gamma_{2} \in \{v(v+1), (v+1)s, ss\}$ common at the locations $\tilde{N} - 2$ and $\tilde{N}-1$ and only focus on its effect in $\frac{\tilde{P}_{cs,\tilde{N}}^{-1}B_{\tilde{N}}^{'}-B_{\tilde{N}}}{\tilde{P}_{cs,\tilde{N}-1}^{\tilde{q}_{cs,\tilde{N}-1}}}$. To handle the former case, the term $\beta_{01,\gamma_{2}E}$ from $B_{\tilde{N}}^{'}$ is of the form $\sum_{i=1}^{\tilde{q}_{cs,\tilde{N}-1}+ 1}\tilde{P}_{cs,\tilde{N}-1}^{\tilde{q}_{cs,\tilde{N}-1}+1-i}\tilde{P}_{cs,\tilde{N}}^{i-1}$, whereas the term $\beta_{01,\gamma_{2}E}$ from $B_{\tilde{N}}$ is of the form $\sum_{i=1}^{\tilde{q}_{cs,\tilde{N}-1}}\tilde{P}_{cs,\tilde{N}-1}^{\tilde{q}_{cs,\tilde{N}-1}-i}\tilde{P}_{cs,\tilde{N}}^{i-1}$. By taking $\tilde{P}_{cs, \tilde{N}-1}^{\tilde{q}_{cs,\tilde{N}-1}}$ common from both the above terms, the difference of the two corresponding terms in $\frac{\tilde{P}_{cs,\tilde{N}}^{-1}B_{\tilde{N}}^{'}-B_{\tilde{N}}}{\tilde{P}_{cs,\tilde{N}-1}^{\tilde{q}_{cs,\tilde{N}-1}}}$ is $\frac{1}{\tilde{P}_{cs,\tilde{N}}}$, and this is because of the equality 
		\begin{eqnarray}
			\label{PDF_cluster_semi_cumm_3}
			\sum_{i=1}^{\tilde{q}_{cs,\tilde{N}-1}}\tilde{P}_{cs,\tilde{N}-1}^{-i}\tilde{P}_{cs,\tilde{N}}^{i-1} - \sum_{i=1}^{\tilde{q}_{cs,\tilde{N}-1}+1}\tilde{P}_{cs,\tilde{N}-1}^{1-i}\tilde{P}_{cs,\tilde{N}}^{i-2} = - \frac{1}{\tilde{P}_{cs,\tilde{N}}}.
		\end{eqnarray}
		This completes the proof that $\frac{\tilde{P}_{N}^{-1}B_{\tilde{N}}^{'}-B_{\tilde{N}}}{\tilde{P}_{cs,\tilde{N}-1}^{\tilde{q}_{cs,\tilde{N}-1}}}$ does not contain $\tilde{q}_{cs,\tilde{N}-1}$ in it from sequences ending with `01'. Note that this argument holds when $\gamma_{2} \in \{(v+1)s, ss\}$. To handle the sequences that end with `10', we can have three types of terms based on the positions of node $v$ and node $v+1$ and their status of borrowing the residual ARQs. One type of term is $\beta_{01,ssE}$ which contributes to $B_{\tilde{N}}^{'}$ a term of the form $\sum_{i=1}^{\tilde{q}_{cs,\tilde{N}-2}} \tilde{P}_{cs,\tilde{N}-2}^{\tilde{q}_{cs,\tilde{N}-2}-i}\sum_{k=0}^{i-2}\tilde{P}_{cs,\tilde{N}-1}^{\tilde{q}_{cs,\tilde{N}-1}+ 1 + k}$. Similarly, the term $\beta_{01,ssE}$ contributes to $B_{\tilde{N}}$ a term of the form $\sum_{i=1}^{q_{cs,\tilde{N}-2}} \tilde{P}_{cs,\tilde{N}-2}^{\tilde{q}_{cs,\tilde{N}-2}-i} \sum_{k=0}^{i-2}P_{cs,\tilde{N}-1}^{\tilde{q}_{\tilde{N}-1} + k}$. After taking out $\tilde{P}_{cs,\tilde{N}-1}^{\tilde{q}_{\tilde{N}-1}}$ common, we can evaluate that $\frac{\tilde{P}_{cs,\tilde{N}}^{-1}B_{\tilde{N}}^{'}-B_{\tilde{N}}}{\tilde{P}_{cs,\tilde{N}-1}^{q_{\tilde{N}-1}}}$ does not contain $\tilde{q}_{\tilde{N}-1}$.
		Furthermore, the other type of term is $\beta_{01,v(v+1)E}$ which contributing to $B_{\tilde{N}}^{'}$ a term of the form $\sum_{i=1}^{\tilde{q}_{cs,\tilde{N}-2}}psp_{cs,\tilde{N}-2}(\tilde{q}_{cs,\tilde{N}-2}-i+1) \sum_{j=1}^{i-1}\tilde{P}_{cs,\tilde{N}-1}^{\tilde{q}_{cs,\tilde{N}-1}+j}$. Similarly, the term $\beta_{01,v(v+1)E}$ contributing to $B_{\tilde{N}}$ is of the form $\sum_{i=1}^{\tilde{q}_{cs,\tilde{N}-2}} psp_{cs,\tilde{N}-2}(\tilde{q}_{cs,\tilde{N}-2}-i+1) \sum_{j=1}^{i-1}\tilde{P}_{cs,\tilde{N}-1}^{\tilde{q}_{\tilde{N}-1}+j-1}$. Therefore, after taking out $\tilde{P}_{cs,\tilde{N}-1}^{\tilde{q}_{\tilde{N}-1}}$ common, we can show that $\frac{\tilde{P}_{cs,\tilde{N}}^{-1}B_{\tilde{N}}^{'}-B_{\tilde{N}}}{\tilde{P}_{cs,\tilde{N}-1}^{\tilde{q}_{\tilde{N}-1}}}$ does not contain $\tilde{q}_{\tilde{N}-1}$. Similar result can also be proved for the third type of term $\beta_{11,v(v+1)E}$. This completes the proof that $\frac{B_{\tilde{N}}^{'}-\tilde{P}_{cs,\tilde{N}}B_{\tilde{N}}}{\tilde{P}_{cs,\tilde{N}-1}^{\tilde{q}_{cs,\tilde{N}-1}}}$ does not contain $\tilde{q}_{cs,\tilde{N}-1}$ in it from sequences ending with `10'.
		
		Henceforth, equation \eqref{eq:indept_binary} is written as $\tilde{R}_{1,\tilde{N}} = \tilde{P}_{cs,\tilde{N}}^{\tilde{q}_{cs,\tilde{N}}}\tilde{R}_{2,\tilde{N}}$, wherein $\tilde{R}_{1,\tilde{N}} \triangleq \bigg(\frac{B_{\tilde{N}-1}}{\prod_{i=1, i \neq j, s.t. \tilde{P}_{cs, j} = P_{v}}^{\tilde{N}-2}\tilde{P}_{cs,i}^{q_{cs,i}}} \bigg)$ and $\tilde{R}_{2, N} \triangleq  \frac{(\tilde{P}_{cs,\tilde{N}}^{-1}B_{\tilde{N}}^{'}-B_{\tilde{N}})}{\prod_{i=1, i \neq j, s.t. \tilde{P}_{cs, j} = P_{v}}^{\tilde{N}-1}\tilde{P}_{cs,i}^{\tilde{q}_{cs,i}}}$ do not contain the terms $\tilde{P}_{cs,\tilde{N}}^{\tilde{q}_{cs,\tilde{N}}}$ and $\tilde{q}_{cs,\tilde{N}-1}$. Hence, $\tilde{R}_{1,\tilde{N}}$ and $\tilde{R}_{2,\tilde{N}}$ are constants since $\{\tilde{P}_{cs,i} ~|~i = 1,2,\dots,\tilde{N}\}$ and $\{\tilde{q}_{cs,i} ~|~ i = 1,2,\dots,\tilde{N}-2\}$ are fixed. Now, we can rewrite the equality condition as $\tilde{P}_{\tilde{N}}^{\tilde{q}_{\tilde{N}}}= \frac{\tilde{R}_{1,\tilde{N}}}{\tilde{R}_{2,\tilde{N}}}$, or as $\tilde{q}_{cs,\tilde{N}} = \frac{\big(\log \frac{\tilde{R}_{1,\tilde{N}}}{\tilde{R}_{2,\tilde{N}}}\big)}{\log \tilde{P}_{cs,\tilde{N}}}$. Note that in our work, we have a condition that $q_{cs,i} \in \mathbb{Z}_{+} $, however, the solution of $\tilde{q}_{cs,\tilde{N}}= \frac{\big(\log \frac{\tilde{R}_{1,\tilde{N}}}{\tilde{R}_{2,\tilde{N}}}\big)}{\log \tilde{P}_{cs,\tilde{N}}}$ may not belong to $\mathbb{Z}_{+}$. It implies that to find the optimal solution which lies in $\mathbb{Z}_{+}$, we need to obtain either $\ceil{\tilde{q}_{cs,\tilde{N}}}$ or $\floor{\tilde{q}_{cs,\tilde{N}}}$ from the equality condition. It can be observed that $\ceil{\tilde{q}_{cs,\tilde{N}}}$ will decrease $\tilde{P}_{cs,\tilde{N}}^{\ceil{q_{cs,\tilde{N}}}}$, and this implies that $pdp_{cs,\tilde{N}} > pdp_{cs,\tilde{N}}^{'}$, and this is a sub-optimal solution because when we give one more ARQ from the last hop to the second last hop, PDP decreases. On the other hand, if we use $\floor{\tilde{q}_{cs,\tilde{N}}}$, then $\tilde{P}_{cs,\tilde{N}}^{\floor{\tilde{q}_{cs,\tilde{N}}}}$ increases, which implies $pdp_{cs,\tilde{N}} < pdp_{cs,\tilde{N}}^{'}$. Therefore, on giving one more ARQ from the last hop to second last hop, PDP increases, and this implies that using $\tilde{q}_{cs,\tilde{N}}= \lfloor \frac{\big(\log \frac{\tilde{R}_{1,\tilde{N}}}{\tilde{R}_{2,\tilde{N}}}\big)}{\log \tilde{P}_{cs, \tilde{N}}} \rfloor$ in $\tilde{\mathbf{q}}_{cs,\tilde{N}}$ captures the optimal solution conditioned on the first $\tilde{N}-2$ ARQ numbers. Thus, on fixing $\tilde{q}_{1},\tilde{q}_{2},\dots,\tilde{q}_{\tilde{N}-2}$, we can analytically compute $\tilde{q}_{\tilde{N}}$, and also compute $\tilde{q}_{\tilde{N}-1}$ using the relation $\tilde{q}_{cs,\tilde{N}-1} = q_{sum} - \sum_{t = 1, t \neq \tilde{N}-1}^{\tilde{N}} \tilde{q}_{cs,t}$. 
	\end{IEEEproof}

	Henceforth, the reduction technique in Theorem \ref{thm:complexity_reduction_cs_case1_2} is referred to as the one-fold technique since the search space for the $\tilde{N}$-hop network is reduced to that of an $(\tilde{N}-2)$-hop network. In the next section, we present the results on complexity reduction for Case-3.
	\subsection{Complexity Reduction for Case-3}	
	\label{prop_cluseter_at_last}
	When the cluster is placed at the last position in the network, it is not possible to obtain the results along the lines of Theorem \ref{thm:complexity_reduction_cs_case1_2}. This is because the PDP at node $v$ is not of the form of the PDP of a semi-cumulative node. To circumvent this problem, we present a new one-fold algorithm, which is as explained below. First, we split the $N$-hop network into two parts namely $\hat{N}_{1}$-hop and $\hat{N}_{2}$-hop sub-networks such that $\hat{N}_{1}= N_{s1} + 1$ includes all the nodes of the semi-cumulative network along with the first node of the cluster, and $\hat{N}_{2}= N_{c2}-1$ includes all the nodes of the cluster except the first node. Let the outage probabilities and ARQ vector of $\hat{N}_{1}$-hop network be given by $\mathbf{\hat{P}}_{1} = [\hat{P}_{1},\hat{P}_{2}, \ldots, \hat{P}_{ \hat{N}_{1}}]= [P_{s1,1},P_{s1,2}, \ldots, P_{s1,N_{s1}}, P_{c2,1}]$, where $\hat{P}_{ \hat{N}_{1}} = P_{c2,1}$ (outage probability of the first node of the cluster) and $\mathbf{\hat{q}}_{1} = [\hat{q}_{1},\hat{q}_{2}, \ldots, \hat{q}_{ \hat{N}_{1}}] = [q_{s1,1}, q_{s1, 2}, \ldots, q_{s1, N_{s1}}, q_{c2,1}]$. Since the $\hat{N}_{1}$-hop network is also semi-cumulative, we apply Theorem \ref{complexity_them_1} on it by feeding a total of $q_{sum,\hat{N}_{1}} = q_{sum}-(N_{c2}-1)$ ARQs. This way, we compute the list of ARQ distributions for the $\hat{N}_{1}$-hop network, which is as large as the search space for $(\hat{N}_{1}-2)$-hop network. Subsequently, for each ARQ distribution in the list, we transfer $(N_{c2}-1)$ ARQs to $q_{\hat{N}_{1}}$ by assigning $q_{c2, j} = 0$ for $2 \leq j \leq N_{c2}$, and then compute the optimal ARQ distribution of the $\hat{N}$-hop network. Through this process, the size of the list is reduced to that of a $(\hat{N}_{1}-2)$-hop network thereby reducing the complexity. The pseudocode for the proposed method is provided in Algorithm \ref{algo:case3}.
	\begin{algorithm}
		\caption{\label{algo:case1} Multi-folding list algorithm for Case-1 of the CSC strategy}
		\begin{algorithmic}[1]
			\begin{small}
				\Require $N$, $N_{c1}$, $\tilde{N}$, $q_{sum}$, Outage probabilities of the links
				\Ensure $\mathcal{L}_{final} $ - List of ARQ distributions in search space
				\State $\mathcal{L}_{k} = \{\phi\}$ for $ k= 1,2,\ldots,\tilde{N}$ 
				\If {$\tilde{N}$ is odd} \Comment{Start with fixing $\tilde{q}_{cs,1}$ and $\tilde{q}_{cs,2}$}
				\State $\mathcal{L}_{1} = \{ [N_{c1}-1, q_{sum}-(\tilde{N}-1) +1]\}$ 
				\State Assign $p=3$
				\For {$j=p:2:\tilde{N}$}
				\For {$i_{1}=1:|\mathcal{L}_{j-2}|$}
				\State $[\tilde{q}_{cs,1}, \ldots, \tilde{q}_{cs,j-2}] = \mathcal{L}_{j-2}({i_{1}})$
				\State Compute $\tilde{q}_{cs,j}$ from $[\tilde{q}_{cs,1}, \ldots, \tilde{q}_{cs,j-2}]$ by using Theorem \ref{thm:complexity_reduction_cs_case1_2}
				\For {$\tilde{q}_{sum,j} = (j+(N_{c1}-1)-1): (q_{sum}-(\tilde{N}-j)+1)$}.
				\State Compute $\tilde{q}_{cs,j-1} = \tilde{q}_{sum,j} - \sum_{t=1, t\neq j-1}^{j}\tilde{q}_{cs,t}$.
				\State Insert $[\mathcal{L}_{j-2}(i_{1})||\tilde{q}_{cs,j-1}||\tilde{q}_{cs,j}]$ to $\mathcal{L}_{j}$ if $\tilde{q}_{cs,j-1} \geq 0$
				\EndFor
				\EndFor
				\EndFor
				\State $\mathcal{L}_{final}= \mathcal{L}_{\tilde{N}}$.
				\ElsIf{$\tilde{N}$ is even} \Comment{Start with fixing $\tilde{q}_{cs,1}$ and $\tilde{q}_{cs,2}$}
				\State $\mathcal{L}_{2} = \{\{\tilde{q}_{cs,1},\tilde{q}_{cs,2}\} \in \mathbb{Z}_{+}^{2} | \tilde{q}_{cs,1}+\tilde{q}_{cs,2} \in [N_{c1}, q_{sum}-(\tilde{N}-2)+1 ]\}$
				\State Assign $p=4$, and repeat steps from line number $5$ to $15$
				\EndIf
			\end{small}
		\end{algorithmic}
	\end{algorithm}	  
\section{Low-Complexity Algorithms for Cluster Based Semi-Cumulative Strategy}
For large values of $\tilde{N}$, the one-fold technique might not be feasible to implement in practice. Therefore, to further reduce the complexity, we propose multi-folding and greedy algorithms for all the three cases. 
\subsection{List Generation using Multi-Folding for the CSC Strategy}
\label{sec:multi-algo_cs}
In the proposed multi-folding algorithm, instead of folding the network once from $\tilde{N}$-hop to $(\tilde{N}-2)$-hop, we fold it multiple times to $(\tilde{N}-4)$-hop, $(\tilde{N}-6)$-hop and so on, upto a $2$-hop network or a $1$-hop network depending on $\tilde{N}$, and the positions of node $v$ and node $v+1$ in the network. The pseudocodes of the multi-folding algorithm are presented in Algorithm \ref{algo:case1} for Case-1,  Algorithm \ref{algo:case3} for Case-3, and Algorithm \ref{algo:case2} for Case-2. In Case-1, we use Algorithm \ref{algo:case1} with $\tilde{N}=N-(N_{c2}-1)+1$. Similar to Algorithm \ref{algo:sc}, for the first $j$-hop network, such that $j \in \{3, 4\}$, we fix a total number of ARQs, denoted by $q_{sum, j}$, in the range $[(j+N_{c1}-2), q_{sum,\tilde{N}}-(\tilde{N}-j)+1]$, and then compute $q_{j-1}$ and $q_{j}$ using the ratio $\tilde{R}_{j}$ from Theorem \ref{thm:complexity_reduction_cs_case1_2}. Subsequently, using the list from the $j$-hop network, a list of ARQs is obtained for the $j+2$-hop network, eventually generating a list for the $\tilde{N}$-hop network. Similarly, in Case-3, we use Algorithm \ref{algo:case3} with $\tilde{N}=N-(N_{c2}-1)$. The pseudocode presented in the algorithm captures the ideas presented in Section \ref{prop_cluseter_at_last}. In this case, the multi-folding algorithm as discussed in Section \ref{sec:proposed_method} is applicable on the $\hat{N}_{1}$-hop network which only comprises semi-cumulative nodes. Finally, in Case-2, we use Algorithm \ref{algo:case2} with $\tilde{N}=N-(N_{c2}-1)+1$. Here, when the network is reduced (or folded) to a $j$-hop network, we need to identify the locations of node $v$ and node $v+1$ in the folded network. This is because the ARQs for node $v$ and node $v+1$ must not be computed in the same iteration as it does not result in complexity reduction. Therefore, we split the virtual $\tilde{N}$-hop network into three parts, namely: $\tilde{N_{1}}$-hop, $\tilde{N_{2}}$-hop and $\tilde{N_{3}}$-hop networks such that $\tilde{N}= \tilde{N_{1}}+\tilde{N_{2}}+\tilde{N_{3}}$, wherein $(\tilde{N_{1}}+\tilde{N_{2}})$-hop networks contain the semi-cumulative nodes preceding the cluster along with node $v$, and $\tilde{N_{3}}$-hop network contains node $v+1$ along with the semi-cumulative nodes that follow the cluster. Note that this case invokes results from Algorithm \ref{algo:case3} while folding within the $\tilde{N_{1}}$-hop network.    
	\begin{algorithm}
		\caption{\label{algo:case3} Multi-fold algorithm for Case-3 of the CSC strategy}
		\begin{algorithmic}[1]
			\begin{small}
				\Require $\hat{N}_{1}$, $\hat{N}_{2}$, $q_{sum}$, $q_{sum,\hat{N}_{1}} =q_{sum}-(N_{c2}-1)$, $\mathbf{\hat{P}}_{1} = [\hat{P}_{1},\hat{P}_{2}, \ldots, \hat{P}_{ \hat{N}_{1}}]$
				\Ensure $\mathcal{L}_{final} $ - List of ARQ distributions in search space
				\If {$\hat{N}_{1}$ is odd}   
				\State $\mathcal{L}_{1} = \{ [ 1, q_{sum,\hat{N}_{1}}-(\hat{N}_{1}-1)+1  ]\}$
				\State Assign $p=3$
				\For {$j=p:2:\hat{N}_{1}$}
				\For {$i_{1}=1:|\mathcal{L}_{j-2}|$}
				\State $[\hat{q}_{1}, \ldots, \hat{q}_{j-2}] = \mathcal{L}_{j-2}({i_{1}})$
				\State Compute $\hat{q}_{j}$ from $[\hat{q}_{1}, \ldots, \hat{q}_{j-2}]$ using Theorem \ref{complexity_them_1}
				\For {$\tilde{q}_{sum,j} = j: (q_{sum, \hat{N_{1}}}-(\hat{N}_{1}-j)+1)$}
				\State Compute $\hat{q}_{j-1} = \tilde{q}_{sum,j} - \sum_{t=1, t\neq j-1}^{j}\hat{q}_{t}$
				\State Insert $[\mathcal{L}_{j-2}(i_{1})||\hat{q}_{j-1}||\hat{q}_{j}$] in $\mathcal{L}_{j}$ if $\hat{q}_{j-1} \geq 0$ \&\& $j<\hat{N}_{1}$
				\State Insert $[\mathcal{L}_{j-2}(i_{1})||\hat{q}_{j-1}||\hat{q}_{j} + \hat{N}_{2}$] in $\mathcal{L}_{j}$ if $\hat{q}_{j-1} \geq 0$ \&\& $j =\hat{N}_{1}$
				\EndFor
				\EndFor
				\EndFor
				\State $\mathcal{L}_{final}= \mathcal{L}_{\hat{N}_{1}}$.
				\ElsIf{$\hat{N}_{1}$ is even}
				\Comment{Start with fixing $\hat{q}_{1}$ and $\hat{q}_{2}$}
				\State $\mathcal{L}_{2} = \{\{\hat{q}_{1},\hat{q}_{2}\} \in \mathbb{Z}_{+}^{2} | \hat{q}_{1}+\hat{q}_{2} \in [2, q_{sum,\hat{N}_{1}}-(\hat{N}-2)+1 ]\}$
				\State Assign $p=4$, and repeat steps from line number $4$ to $15$
				\EndIf
			\end{small}
		\end{algorithmic}
	\end{algorithm}	
	\begin{algorithm}
		\caption{\label{algo:case2} Multi-folding list algorithm for Case-2 of the CSC strategy}
		\begin{algorithmic}[1]
			\begin{small}
				\Require $N$, $N_{s1}$, $N_{c2}$, $N_{s3}$, $\tilde{N}$, $q_{sum}$, $\tilde{\mathbf{P}}_{cs} = [\tilde{P}_{cs,1}, \tilde{P}_{cs,2}, \ldots, \tilde{P}_{cs,\tilde{N}}]$
				\Ensure $\mathcal{L}_{final} $ - List of ARQ distributions in search space
				\State $\mathcal{L}_{k} = \{\phi\}$ $k = 1,2,\ldots,\tilde{N}$
				\State Split network $\tilde{N}= \tilde{N}_{1}+\tilde{N}_{2}+\tilde{N}_{3}$ such that $\tilde{N}_{1} = N_{s1}+1$, $\tilde{N}_{2} = N_{c2}-2$ and $\tilde{N}_{3} = N_{s3}+1$
				\State Assign $\mathbf{q}_{sum, \tilde{N_{1}}} = [N_{s1}+1, q_{sum}-(\tilde{N_{2}}+\tilde{N}_{3})]$
				\For{$k_{1}= 1:|\mathbf{q}_{sum, \tilde{N_{1}}}|$}
				\State Call Algorithm \ref{algo:case3} with $q_{sum,\hat{N_{1}}}= \mathbf{q}_{sum, \tilde{N_{1}}}(k_{1})$, $\hat{N}_{1} = \tilde{N_{1}}$, $\hat{\mathbf{P}}_{1} = \tilde{\mathbf{P}}_{\tilde{N}_{1}}$ and $\hat{N}_{2} = \tilde{N}_{2}$
				\State $\mathcal{L}_{\tilde{N}_{1},k_{1}}= \{\mathcal{L}_{\hat{N}_{1}} | \tilde{q}_{j+1} \neq 0\ \text{for} \ \tilde{q}_{j} = 0 \ \text{or} \ 1 \ \text{where} \ j \in \{1,2,\ldots, \tilde{N}_{1}-2 \}$ and $\tilde{q}_{\tilde{N}_{1}} \geq N_{c2}-1$ for $\tilde{q}_{\tilde{N}_{1}-1} \leq 1 $ \}
				\State Insert $\mathcal{L}_{\tilde{N}_{1},k_{1}}$ into $\mathcal{L}_{\tilde{N}_{1}}$
				\EndFor
				\State Assign $p=\tilde{N}_{1}+2$
				\State Assign $\bar{N}= \tilde{N}-1$ if $\tilde{N}_{3}$ is odd, otherwise, assign $\bar{N}= \tilde{N}$ 
				\For {$j=p:2:\bar{N}$}
				\For {$i_{1}=1:|\mathcal{L}_{j-2}|$}
				\State $[\tilde{q}_{1}, \ldots, \tilde{q}_{j-2}] = \mathcal{L}_{j-2}({i_{1}})$
				\State Compute $\tilde{q}_{j}$ from $[\tilde{q}_{1}, \ldots, \tilde{q}_{j-2}]$ using Theorem \ref{thm:complexity_reduction_cs_case1_2}
				\For {$\tilde{q}_{sum,j} = j+\tilde{N_{2}}: (q_{sum}-(\tilde{N}-j)+1)$}. 
				\State Compute $\tilde{q}_{j-1} = \tilde{q}_{sum,j} - \sum_{t=1, t\neq j-1}^{j}\tilde{q}_{t}$.
				\State Insert $[\mathcal{L}_{j-2}(i_{1})||\tilde{q}_{j-1}||\tilde{q}_{j}]$ in $\mathcal{L}_{j}$ if $\tilde{q}_{j-1} \geq 0$
				\EndFor
				\EndFor
				\EndFor
				\If {$\tilde{N}_{3}$ is odd}
				\For {$j_{1}= 1:|\mathcal{L}_{\tilde{N}-1}|$} 
				\State Compute $\tilde{q}_{\tilde{N}} = q_{sum} - sum(\mathcal{L}_{\tilde{N}-1}(j_{1}))$, where $sum(\cdot)$ is the sum of elements in $\mathcal{L}_{\tilde{N}-1}(j_{1})$
				\State Insert $[\mathcal{L}_{\tilde{N}-1}(j_{1})||\tilde{q}_{\tilde{N}}]$ in $\mathcal{L}_{\tilde{N}}$ if $\tilde{q}_{\tilde{N}} \geq 0$ 
				\EndFor
				\EndIf
			\end{small}
			\State \begin{small}$\mathcal{L}_{final}= \mathcal{L}_{\tilde{N}}$\end{small}
	 \end{algorithmic}
	\end{algorithm}
	\subsection{Multi-Folding based Greedy Algorithms for the CSC Strategy}
	\label{sec:greedy-algo_cs}
	To further reduce the size of the search space from that in Algorithms \ref{algo:case1}, \ref{algo:case3} and \ref{algo:case2}, we propose to  retain the ARQ distribution that gives the minimum PDP for a given $\tilde{q}_{sum,j}$ from the list $\mathcal{L}_{j}$. This way, only one ARQ distribution survives for a given $\tilde{q}_{sum,j}$, thereby significantly reducing the list size when the algorithm traverses to $\tilde{q}_{sum,\hat{N}_{1}}$ (refer to  Algorithms \ref{algo:case3}, \ref{algo:case2})  and $\tilde{q}_{sum,\tilde{N}_{1}}$ (refer to Algorithm \ref{algo:case1}). In addition, for the ARQ distribution that survives for a given $\tilde{q}_{sum,j}$, we generate one more ARQ distribution by giving one ARQ from the last node to the penultimate node for every $j$. This is along the similar lines of the greedy algorithm for the multi-folding algorithm in the SC strategy. If we capture the number of computations required to arrive at the final lists, it is clear that the greedy algorithm offers minimum complexity owing to fewer surviving distributions at each level. 
	\section{Simulation Results on Cluster Based Semi-Cumulative Strategy}
	\label{sec:cs_simulations}
	In this section, we present simulation results to showcase the benefits of using a cluster inside the $N$-hop semi-cumulative network for Case-1, Case-2 and Case-3. First, we present simulation results on the PDP and the complexity reduction of the proposed algorithms. For the experiment set up, we consider a $6$-hop network such that Case-1 has $N_{c1}=3$ and $N_{s2}=3$. Similarly, in Case-2, we have $N_{s1}=2$, $N_{c2}=3$ and $N_{s3}=1$, and in Case-3, we have $N_{s1}=3$ and $N_{c2}=3$. Similar to Section \ref{sec:Sims}, we use the saddle-point approximation in \cite[Theorem 2]{V_Poor} on \eqref{eq:outage_prob_link} to compute $\{P_{i}, 1 \leq i \leq N\}$, for a given $\mathbf{c}$ and $\mbox{SNR}$, and also use $K = 500$.
	\begin{figure}
		\begin{center}
			\includegraphics[scale=0.55]{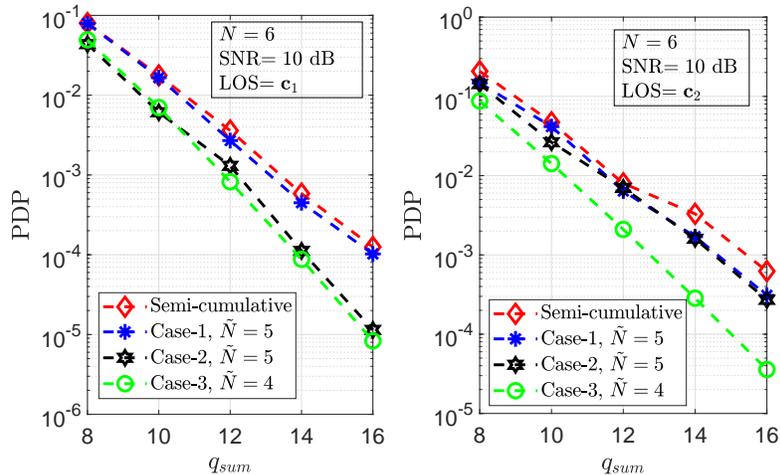}
			\vspace{-0.65cm}
			\caption{\label{fig:PDP_allCases_c1c2_10}PDP comparison: Case-1 (first three hops constitute cluster), Case-2 (third hop to fifth hop forming a cluster) and Case-3 (fourth hop to sixth hop forming a cluster) with $\mathbf{c}_{1}=[0.9,0.2,0.4,0.7,0.1,0.5]$ and $\mathbf{c}_{2}=[0.3,0.3,0.3,0.3,0.3,0.3]$ at rate $R=1$.}
		\end{center}
	\end{figure}
	\begin{figure*}[h]
		\begin{center}
			\includegraphics[scale=0.45]{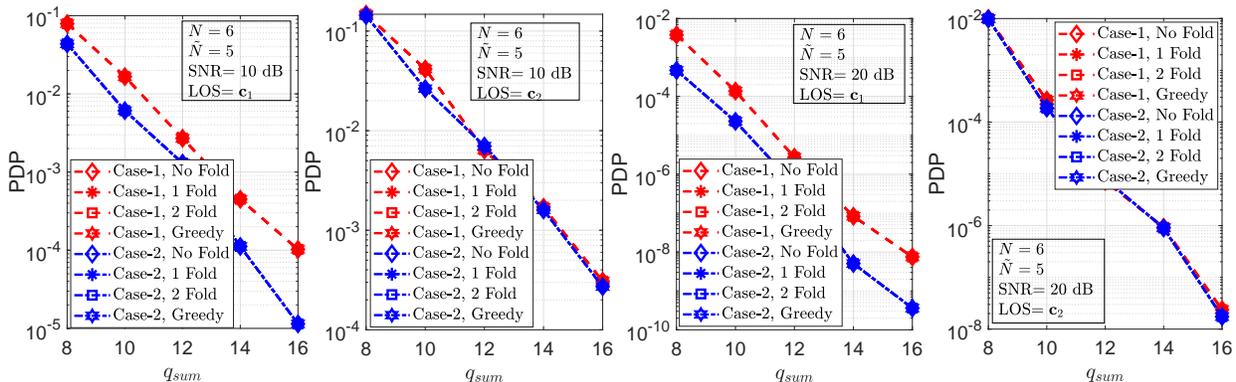}
			\vspace{-0.4cm}
			\caption{\label{fig:PDP_Cases12_c1c2_10_20}Plots depicting the PDP comparison for Case-1 (first three hops constitute cluster) and Case-2 (third hop to fifth hop forming a cluster) with  no-fold, 1-fold, 2-fold and greedy strategies where $\mathbf{c}_{1}=[0.9,0.2,0.4,0.7,0.1,0.5]$ and $\mathbf{c}_{2}=[0.3,0.3,0.3,0.3,0.3,0.3]$.}
		\end{center}
	\end{figure*}
	
\begin{figure*}[]
		\begin{center}
			\includegraphics[scale=0.40]{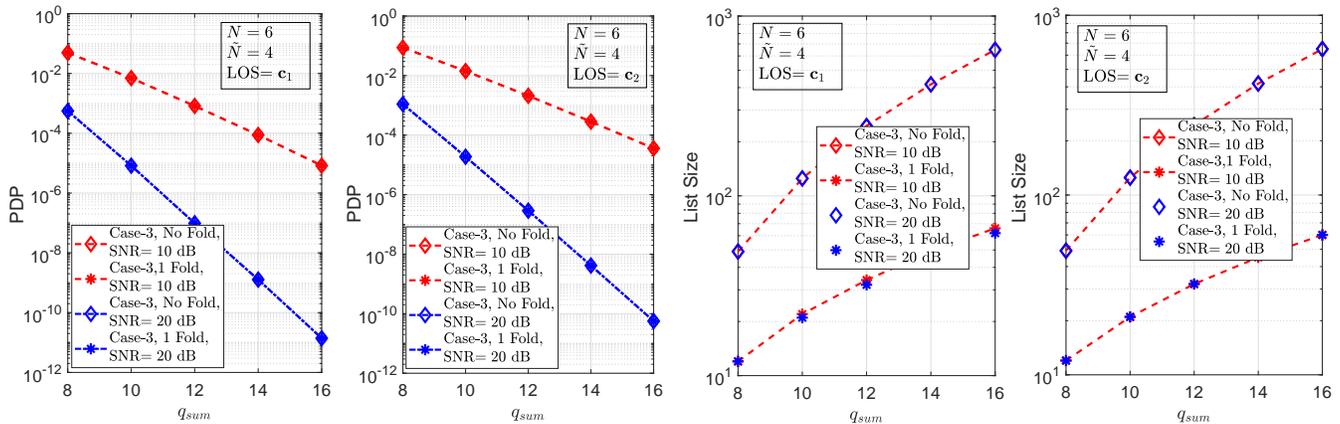}
			\vspace{-0.4cm}
			\caption{\label{fig:PDP_Case3_10_20} On the left: Plots illustrating the PDP comparison for Case-3 with no-fold, 1-fold algorithm where $\mathbf{c}_{1}=[0.9,0.2,0.4,0.7,0.1,0.5]$ and $\mathbf{c}_{2}=[0.3,0.3,0.3,0.3,0.3,0.3]$, SNR = $10$ dB, rate $R=1$. In this case, the last three hops constitute the cluster. On the right: Plots depicting the reduction in list size for Case-3 with exhaustive search, and 1-fold algorithms.}
		\end{center}
	\end{figure*}

	In Fig. \ref{fig:PDP_allCases_c1c2_10}, we plot the minimum PDP offered by the optimal ARQ distribution for each case as a function of $q_{sum}$. From Fig. \ref{fig:PDP_allCases_c1c2_10}, it is clear that Case-3 shows a great improvement in PDP because the number of nodes contributing to node $v$ is three, whereas in Case-1 and Case-2, we have node $v+1$, such that if it uses residual ARQs from its previous node, then, the next node in the chain cannot make use of residual ARQs from node $v+1$. To showcase the significance of multi-folding and the greedy algorithms for Case-1 and Case-2, in Fig. \ref{fig:PDP_Cases12_c1c2_10_20}, we plot the minimum PDP from the list of ARQ distributions by making use of exhaustive search, one-fold, two-fold and greedy algorithms. We use the same LOS vectors as in Fig. \ref{fig:PDP_allCases_c1c2_10}, i.e., $\mathbf{c}_{1}$ and $\mathbf{c}_{2}$ at SNR = $10$ dB. Similarly for Case-3, the results on PDP analysis are shown in the left-side of Fig. \ref{fig:PDP_Case3_10_20}. It can be observed that folding techniques provide near-optimal ARQ distributions. 
	
\begin{figure*}[]
		\begin{center}
			\includegraphics[scale=0.44]{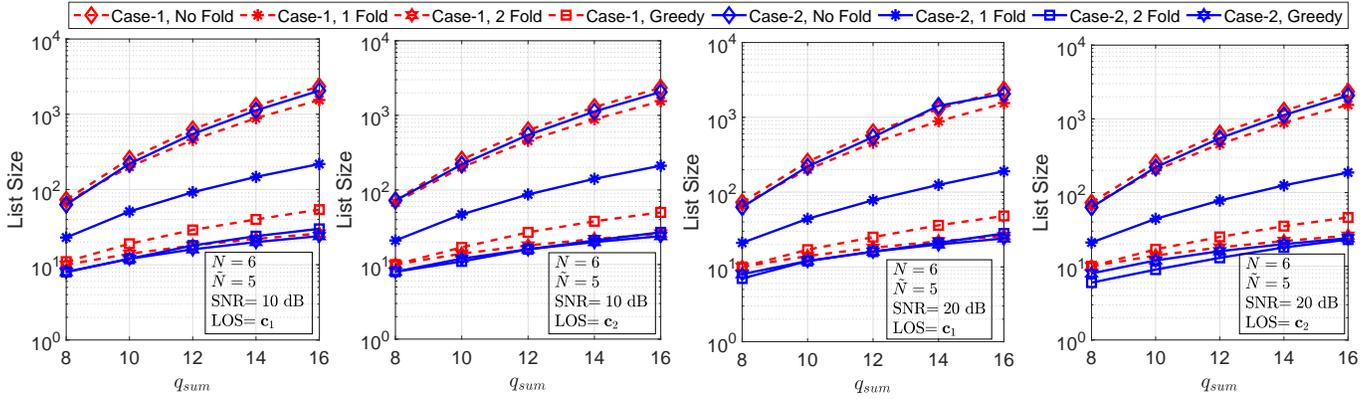}
			\vspace{-0.8cm}
			\caption{\label{fig:list_Cases12_c1c2_10_20}Plots depicting the reduction in list size for Case-1 (first three hops constitute cluster) and Case-2 (third hop to fifth hop forming a cluster) with  no fold, 1 Fold, 2 Fold and Greedy strategies where $\mathbf{c}_{1}=[0.9,0.2,0.4,0.7,0.1,0.5]$ and $\mathbf{c}_{2}=[0.3,0.3,0.3,0.3,0.3,0.3]$.}
		\end{center}
	\end{figure*}

	To evaluate the complexity reduction of our algorithms, we plot the list size for all the three cases as a function of $q_{sum}$ in the right side of Fig. \ref{fig:PDP_Case3_10_20} and in Fig. \ref{fig:list_Cases12_c1c2_10_20}. From the plots, we observe a significant reduction in the list size by using the one-fold method, and further reduction in the list size when using the multi-fold and the greedy algorithms. In Case-3, as $\tilde{N}=4$, we could not apply the multi-folding technique because the network-size does not allow to fold more than once. However, for large $\tilde{N}$, we can apply the multi-folding and the greedy algorithms to reduce the list size to a small number.
	\begin{figure}
		\begin{center}
			\includegraphics[scale=0.58]{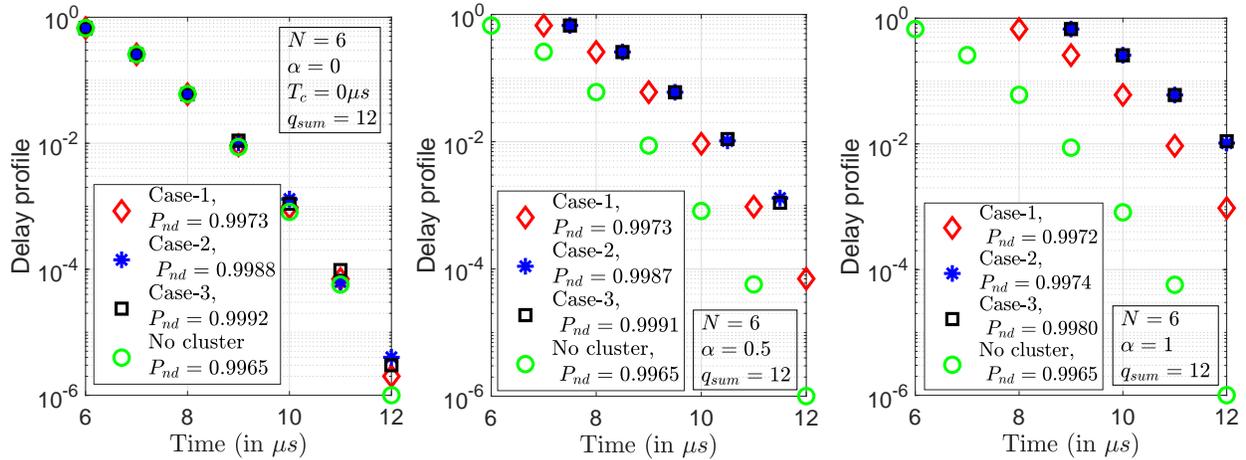}
			\vspace{-0.65cm}
			\caption{\label{fig:delay_profile_10dB} Simulation results on delay profiles (for non-dropped packets) using a 6-hop network with $\mathbf{c}=[0.9,0.2,0.4,0.7,0.1,0.5]$, at rate $R=1$, and $SNR =10 \ dB$  with $10^{6}$ packets (some packets are dropped due to either outage or deadline violations). We have Case-1 (first three hops constitute cluster), Case-2 (third hop to fifth hop constitutes cluster), Case-3 (last three hops constitute cluster) and the SC strategy (with no cluster). In the legend, $P_{nd}$ denotes the fraction of non-dropped packets used to compute the delay profile.}
		\end{center}
	\end{figure}
	
\begin{figure}
	\begin{center}
		\includegraphics[scale=0.6]{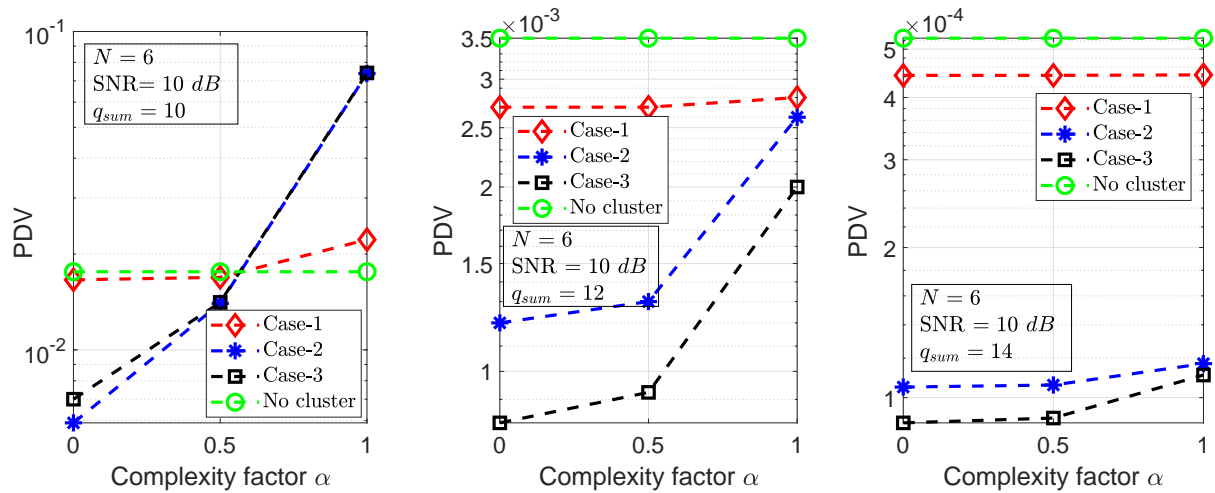}
		\vspace{-0.3cm}
		\caption{\label{fig:PDV_10dB} Simulation results on probability of deadline violation using a 6-hop network with $\mathbf{c} = [0.9,0.2,0.4,0.7,0.1,0.5]$, at rate $R=1$ and SNR = $10 \ dB$ with $10^{6}$ packets. For comparison, we have Case-1, Case-2, Case-3 and the SC strategy (with no cluster). For all the cases, $T_{c}=\alpha T$ microseconds refers to the additional processing delay in updating the counter by a single node.}
	\end{center}
\end{figure}

Owing to the use of ACK/NACK, \cite{our_work_TWC_1} has shown that the average delay offered by the ARQ based strategy is much smaller compared to other strategies for packet retransmissions. We highlight that similar results are also applicable in this work. In the rest of this section, we present delay analysis on packets to study the delay-overhead introduced by the use of the cluster in Case-1, Case-2 and Case-3 scenarios. Let us assume that we want to secure the packets from eavesdropping, and therefore, every node in the cluster encrypts the counter portion of the packet after updating it with the residual ARQs. As a result, when the packet is successfully decoded at the next node, it needs to decrypt it by using an appropriate crypto-primitive. Since this procedure results in an additional processing delay on the packet, we represent this delay by $T_{c}$ microseconds. Assuming that the delay introduced on the packet per hop for each transmission is $T = 1$ microseconds (including both ACK/NACK), we analyse the effect of crypto-primitives on end-to-end delay by choosing $T_{c} = \alpha T$, where $\alpha = 0, 0.5, \mbox{ and }1$. In Fig. \ref{fig:delay_profile_10dB}, we have shown the delay profiles of all three cases along with that of the SC strategy, wherein a delay profile refers to the probability mass function on the delay experienced by an ensemble of packets that reach the destination. Evidently, delay profiles are same for the SC strategy irrespective of the value of $\alpha$. Furthermore, if the effect of $\alpha$ is not considered when designing $q_{sum}$, then there is a non-zero probability that some packets may reach the destination beyond the deadline. Therefore, in addition to PDP, we define a new metric referred to as probability of deadline violation (PDV), which can be defined as the probability that the packets either do not reach the destination before the deadline or get dropped in the network. In Fig. \ref{fig:PDV_10dB}, we plot the PDV of the four strategies as a function of $\alpha$ for a $6$-hop network with different parameters. Since $q_{sum} = 12$, the deadline for packets to reach the destination is $12$ microseconds. The plots confirm that: (i) The PDV of the semi-cumulative network do not change with $\alpha$. (ii) The PDV of the CSC strategy increases with increasing values of $\alpha$; this is because some of the nodes make use of the counter in the packet. (iii) The worst-hit are Case-2 and Case-3 strategies with small $q_{sum}$ as every node in the cluster has to modify the counter, thereby adding a significant delay of $N_{c1}T_{c}$ microseconds to the packet. However, in Case-1, the nodes except the first node of the cluster need to encrypt and decrypt, and hence, the additional delay from the cluster is $(N_{c1}-1)T_{c}$ microseconds. Overall, the simulation results of Fig. \ref{fig:PDV_10dB} show that CSC strategy outperforms the SC strategy when $\alpha$ is small. However, as $\alpha$ increases, the performance degrades. 

\section{Summary}
\label{sec:conclusions}

We have proposed a new family of cooperative ARQ strategies to assist low-latency communication in multi-hop networks. We have derived closed-form expressions on the PDP of the proposed strategies, and have solved the non-linear optimization problem of minimizing their PDP under a sum constraint on the total number of ARQs. We have shown that our strategies outperform the best-known strategies in this space. In this work, we have considered equal processing delays at each node to impose an upper bound on the total number of ARQs. However, in general, when the processing delays at the nodes are unequal, the questions on \emph{How to bound the total number of ARQs?} and subsequently, \emph{How to solve the ARQ optimization problem?} are some interesting directions for future research.


\begin{thebibliography}{16}
	
	\bibitem{Zhou2015}
	Y.~{Zhou}, N.~{Cheng}, N.~{Lu}, and X.~S. {Shen}, ``Multi-UAV-aided networks: Aerial-ground cooperative vehicular networking architecture," \emph{IEEE Vehicular Technology Magazine}, vol.~10, no.~4, pp. 36--44, Dec. 2015.
	
	\bibitem{Shaikh2018}
	F.~S. {Shaikh} and R.~{Wismüller}, ``Routing in multi-hop cellular Device-to-Device (D2D) networks: A survey," \emph{IEEE Communications Surveys Tutorials}, vol.~20, no.~4, pp. 2622--2657, June 2018.
	
	\bibitem{Badarneh2016}
	O.~S. {Badarneh} and F.~S. {Almehmadi}, ``Performance of multihop wireless networks in $\alpha$-$\mu$ fading channels perturbed by an additive generalized Gaussian noise," \emph{IEEE Communications Letters}, vol.~20, no.~5, pp. 986--989, May 2016.
	
	\bibitem{Pocovi2018}
	Pocovi et al., ``Achieving ultra-reliable low-latency communications: Challenges and envisioned system enhancements," \emph{IEEE Network}, vol.~32, no.~2, pp. 8--15, March 2018.
	
	
	\bibitem{Ji2018}
	Ji et al., ``Ultra-reliable and low-latency communications in 5G downlink: Physical layer aspects," \emph{IEEE Wireless Communications}, vol.~25, no.~3, pp. 124--130, June 2018.
	
	\bibitem{Schulz2017}
	P.~{Schulz}, M.~{Matthe}, H.~{Klessig}, M.~{Simsek}, G.~{Fettweis}, J.~{Ansari}, S.~A. {Ashraf}, B.~{Almeroth}, J.~{Voigt}, I.~{Riedel}, A.~{Puschmann}, A.~{Mitschele-Thiel}, M.~{Muller}, T.~{Elste}, and M.~{Windisch}, ``Latency critical IoT applications in 5G: Perspective on the design of radio interface and network architecture," \emph{IEEE Communications Magazine}, vol.~55, no.~2, pp. 70--78, Feb. 2017.
	%
	
	\bibitem{Changyang_She}
	C.~{She}, C.~{Liu}, T.~Q.~S. {Quek}, C.~{Yang}, and Y.~{Li}, ``Ultra-reliable and low-latency communications in unmanned aerial vehicle communication systems," \emph{IEEE Transactions on Communications}, vol.~67, no.~5, pp. 3768--3781, May 2019.
	
	\bibitem{Hong_Ren}
	H.~{Ren}, C.~{Pan}, K.~{Wang}, Y.~{Deng}, M.~{Elkashlan}, and A.~{Nallanathan}, ``Achievable data rate for URLLC-enabled UAV systems with 3-D channel model," \emph{IEEE Wireless Communications Letters}, vol.~8, no.~6, pp. 1587--1590, Dec. 2019.
	
	\bibitem{Chen2018}
	Y.~{Chen}, N.~{Zhao}, Z.~{Ding}, and M.~{Alouini}, ``Multiple UAVs as relays: Multi-hop single link versus multiple dual-hop links," \emph{IEEE Transactions on Wireless Communications}, vol.~17, no.~9, pp. 6348--6359, Sept. 2018.
	
	\bibitem{Mozaffari_LOS}
	M.~{Mozaffari}, W.~{Saad}, M.~{Bennis} and M.~ {Debbah},``Mobile Unmanned Aerial Vehicles (UAVs) for Energy-Efficient Internet of Things Communications," \emph{IEEE Transactions on Wireless Communications}, vol.~16, no.~11, pp. 7574-7589, Nov. 2017. 
	
	\bibitem{Hourani_UAV_application}
	A.~{Al-Hourani}, S.~{Kandeepan}, and A.~{Jamalipour}, ``Modeling air-to-ground path loss for low altitude platforms in urban environments,” in the Proc. of \emph{2014 IEEE Global Communications Conference}, Austin, TX, pp. 2898--2904, Dec. 2014. 
	
	\bibitem{Hourani_UAV_application2}
	A.~{Al-Hourani}, S.~{Kandeepan}, and S.~{Lardner}, ``Optimal lap altitude for maximum coverage,” \emph{IEEE Wireless Commun. Letter}, vol.~3, no.~6, pp. 569--572, Dec. 2014.
	
	\bibitem{Lin_UAV_application}
	X.~{Lin et al.},``The sky is not the limit: LTE for unmanned aerial vehicles,” \emph{IEEE Commun. Mag.}, vol.~56, no.~4, pp. 204--210, Apr. 2018.
	
	\bibitem{Wiemann2005}
	Wiemann et al., ``A novel multi-hop ARQ concept," in \emph{IEEE Vehicular Technology Conference}, pp. 3097--3101, May 2005.	
	\bibitem{our_work_VTC_1}
	Jaya~{Goel} and J.~{Harshan}, ``Minimal Overhead ARQ Sharing Strategies for URLLC in Multi-Hop Networks," 2021 IEEE 93rd Vehicular Technology Conference \emph{VTC2021-Spring}, pp. 1--7, June 2021.
	
	\bibitem{our_work_TWC_1}
	Jaya~{Goel} and J.~{Harshan}, ``Listen to Others’ Failures: Cooperative ARQ Schemes for Low-Latency Communication over Multi-Hop Networks," \emph{IEEE Transactions on Wireless Communications}, vol. 20, no. 09, pp. 6049--6063, Sept. 2021.
	
	
	\bibitem{V_Poor}
	P. Mary et al. ``Finite Blocklength Information Theory: What is the Practical Impact on Wireless Communications?," \emph{IEEE Globecom Workshops (GC Wkshps)}, Washington DC, USA, pp. 1--6, 2016.
	
	\bibitem{polyanskiy}
	Y. Polyanskiy, H. V. Poor and S. Verdu, ``Channel Coding Rate in the Finite Blocklength Regime," in \emph{IEEE Transactions on Information Theory}, vol. 56, no. 5, pp. 2307--2359, May 2010.
	
	\bibitem{MohitSharma2016}
	M.~K.~{Sharma} and C.~R.~{Murthy}, ``Packet drop probability analysis of dual energy harvesting links with retransmission," \emph{IEEE Journal on Selected Areas in Communications}, vol.~34, no.~12, pp. 3646--3660, Dec. 2016.
	
	\bibitem{CoopARQ_EnergyHarvesting}
	M.~{Tacca} and P.~{Monti} and A.~{Fumagalli}, ``Cooperative and reliable ARQ protocols for energy harvesting wireless sensor nodes," \emph{IEEE Transactions on Wireless Communications}, vol.~6, no.~7, pp. 2519-2529, July 2007.
	
	\bibitem{CoopARQ_Diversity}
	L.~{Le} and E.~{Hossain}, ``An analytical model for ARQ cooperative diversity in multi-hop wireless networks," \emph{IEEE Transactions on Wireless Communications}, vol.~7, no.~5, pp. 1786-1791, May 2008.
	
	
	\bibitem{Martin_Serror}
	M.~{Serror}, C.~{Dombrowski}, K.~{Wehrle}, and J.~{Gross}, ``Channel coding versus cooperative ARQ: Reducing outage probability in ultra-low latency wireless communications," in the Proc. of \emph{IEEE Globecom Workshops}, Dec. 2015.
	
	
	
	
	
	
	
	
	
	
	
\end{thebibliography}
\end{document}